\documentclass[a4paper,11pt]{article}
\usepackage{jheppub} 
\usepackage{lineno}
\usepackage{macros}
\usepackage{amsthm}
\usepackage[T1]{fontenc}
\usepackage[utf8]{inputenc}
\DeclareUnicodeCharacter{2010}{-}

\usepackage{color}
\definecolor{dark-gray}{gray}{0.20}
\definecolor{gray}{gray}{0.30}
\definecolor{light-gray}{gray}{0.80}
\definecolor{dark-red}{rgb}{0.7,0,0}
\definecolor{dark-green}{rgb}{0.1,0.4,0}
\definecolor{dark-blue}{rgb}{0.3,0.3,0.7}
\definecolor{dark-blue2}{RGB}{0, 114,178}
\definecolor{light-blue}{rgb}{0.8,0.8,1}
\definecolor{swamp}{RGB}{240, 199, 197}
\definecolor{antiquefuchsia}{rgb}{0.57, 0.36, 0.51}
\newcommand{\bell}{{\boldsymbol{\ell}}}

\usepackage{hyperref}
\usepackage{cleveref}

\hypersetup{
	colorlinks=true,
	linkcolor=dark-blue2,
	citecolor=dark-blue2,
	urlcolor=dark-blue2,
	linktoc=section
}

\usepackage{tocloft}

\begin{document}

\begin{titlepage}
\begin{center}
\rightline{\small }

\vskip 15 mm

{\large \bf

} 
\vskip 11 mm

\begin{center}
{\Large \bfseries Symmetries and M-theory-like Vacua in Four Dimensions}\\[.25em]
\vspace{1cm}
{\bf Shi Chen, Damian van de Heisteeg, Cumrun Vafa}

\vskip 11 mm

{\small \it
 Jefferson Physical Laboratory, \\
Harvard University, Cambridge, MA 02138, USA\\[3mm]
}

\vspace*{1.5em}

\end{center}

\vskip 11 mm

\end{center}
\vskip 17mm

\begin{abstract}
\noindent 
Non-geometric flux vacua have recently been revisited, leading to the remarkable discovery of isolated 4D ${\mathcal N}=1$ supersymmetric Minkowski vacua. These constructions rely on the non-renormalization of the superpotential, which is supported by heuristic arguments. Given the significance of verifying the existence of these isolated M-theory-like vacua, we present alternative symmetry-based arguments that arrive at the same conclusion. Additionally, we leverage these symmetries to argue for the existence of unstable dS solutions as well as supersymmetric AdS solutions.
\end{abstract}

\vfill

\vspace*{13em}


\end{titlepage}

\newpage

\tableofcontents

\section{Introduction}
Construction of reliable string landscape with minimal ($4d,\ {\cal N}=1$) supersymmetry or no supersymmetry is one of the major challenges in constructing realistic models of string theory.  Observations suggest that we have an approximately stable vacuum without supersymmetry and without any massless scalar fields.  Such an example is yet to be constructed from string theory.  Part of the problem is that geometric string constructions generically lead to a plethora of massless fields which control the moduli of compactifications and finding a critical point of such modes is notoriously difficult.  In fact this problem is aggravated, because in weak coupling regimes of the parameter space, where we have analytic control, one can show no such solutions can exist \cite{Dine:1985he}.

This state of affairs forces us to ask whether there are ways to avoid having so many massless moduli and if so find ways to control regions of strong couplings with small or no supersymmetry.  For the former issue, a set of promising ideas that have been pursued include non-geometric string compactifications which typically have less moduli.  These include for example asymmetric orbifolds \cite{Narain:1986qm, Narain:1990mw, Gkountoumis:2023fym, Baykara:2023plc, Gkountoumis:2024dwc,Baykara:2024tjr, Baykara:2025lhl} (and particularly quasi-crystalline compactifications \cite{Harvey:1987da, Baykara:2024vss}), as well as more generally CFT's which add up to the correct central charge and with the right amount of supersymmetry to play the role of a `string background' even though they do not correspond to string propagation on a manifold. Examples of this include some of the Gepner models \cite{Gepner:1987qi} and also certain Landau-Ginzburg (LG) models \cite{Vafa:1988uu,Martinec:1988zu,Greene:1988ut,Vafa:1989xc,Vafa:1991uz,Witten:1993yc, Candelas:1993nd}.  

To address the issue of analytic control in the strong coupling regime not many ideas have been pursued. One idea is to use topological considerations on the total space combined with known boundary behavior at weak coupling to gain some insight \cite{Lust:2024aeg}. The approach we will be following in this paper is to use duality symmetries to gain insight into such strongly-coupled regimes. In particular we focus on isolated points in the light moduli fields with maximal discrete gauge symmetries, required to exist by duality symmetries as well as symmetries of the CFT's, and use these symmetries to constrain the dynamics of both supersymmetric and non-supersymmetric vacua in $4d,\ {\cal N}=1$ supersymmetric theories.  In particular we apply these ideas to the non-geometric LG models initiated in \cite{Becker:2006ks} (see also \cite{Becker:2007dn}) which has been recently revisited in \cite{Ishiguro:2021csu,Bardzell:2022jfh,Becker:2022hse,Becker:2023rqi,  Cremonini:2023suw, Becker:2024ijy, Ishiguro:2024coq} and has led to candidate constructions of isolated $4d,\ {\cal N}=1$ theories without any massless fields \cite{Becker:2024ayh, Rajaguru:2024emw}.\footnote{\color{red} It has recently been pointed out in \cite{Morros:2026yyi} that the tadpole charge in the $(x^4)^{\otimes 6}$ LG model should be 20 instead of the 40 used in these works. In v3 of this paper we have included new unstable dS vacua that saturate this updated bound.  We have not found any Minkowski and AdS vacua which saturate the new bound.  Nevertheless it is plausible that in some other models and other fluxes one could in principle obtain all three types of vacua.  We will keep the changes in the red color for reader's convenience.}  These examples are 4 dimensional analogs of M-theory in 11 dimensions where one has the minimal amount of supersymmetry without any massless scalar fields!  The argument used in these papers to establish such vacua relies heavily on the non-renormalization arguments for the superpotential \cite{Gukov:1999ya} which has only a heuristic justification in the non-geometric case \cite{Becker:2006ks}. Given the importance of establishing such vacua it is important to find alternative arguments. 

A primary goal of this paper is to present alternative arguments based on symmetry principles to argue for the validity of such Minkowski vacua.  In addition we use the same symmetry arguments to argue for the existence of unstable dS vacua, where we are at a saddle point of a positive potential for all scalar fields.  This is in spirit similar to what was done in \cite{Ginsparg:1986wr} in finding critical points of non-supersymmetric $O(16)\times O(16)$ potential by going to a gauge symmetric point where all the massless fields (except for the dilaton) are charged (see \cite{Fraiman:2023cpa} for a recent systematic analysis of such examples). A similar idea was used recently in the context of flux compactification of certain F-theory models to fix all complex structure moduli by focusing on symmetric loci \cite{Grimm:2024fip} (see also \cite{DeWolfe:2004ns, DeWolfe:2005gy, Palti:2007pm, Candelas:2019llw, Kachru:2020sio,Kachru:2020abh, Candelas:2021mwz,Bonisch:2022slo, Candelas:2023yrg, Ducker:2025wfl} for related constructions).   In our models {\it all} moduli are made critical based on symmetry considerations.  Moreover, to our knowledge, we provide the first relatively reliable examples of (unstable) dS solutions in the string landscape. 

The organization of this paper is as follows. In section \ref{sec:nongeometric} we review flux compactifications in non-geometric backgrounds given by LG models. In particular, in subsection \ref{sec:dualitysymmetries} we discuss why duality symmetries such as SL$(2,\mathbb{Z})$ survive the dimensional reduction and orientifold projection. We review the $(x^4)^{\otimes 6}$ and $(x^3)^{\otimes 9}$ LG models in subsections \ref{ssec:x46} and \ref{ssec:x39}. In section \ref{sec:symmetries} we use these symmetries to constrain the expansion of the superpotential and the scalar potential in terms of the fluxes and the complex structure moduli and axio-dilaton. This yields general selection rules even when corrections to the K\"ahler and superpotential are taken into account. In turn, we leverage these selection rules in section \ref{sec:criticalpoints} to argue for critical points of scalar potentials induced by fluxes, such as isolated Minkowski vacua, de Sitter saddle points and AdS vacua. We end the paper with some concluding thoughts in section \ref{sec:conclude}. Two mathematica notebooks are also included as ancillary files detailing the computations in the LG models.

\section{Non-geometric flux compactifications}\label{sec:nongeometric}
In this section we review some general aspects about flux compactifications and LG orbifold models without K\"ahler moduli. We begin in subsection \ref{ssec:fluxcomp} with a general review of Type IIB Calabi--Yau orientifold compactifications with three-form fluxes. In subsection \ref{sec:dualitysymmetries} we then explain how SL$(2,\mathbb{Z})$ duality symmetry descends from 10d Type IIB string theory down to the 4d $\mathcal{N}=1$ supergravity theory. Thereafter we specialize our discussion to LG orbifold models: we explain their advantages as non-geometric backgrounds for 4d $\mathcal{N}=1$ supergravities in subsection \ref{ssec:nongeom}, and review the details of the $(x^4)^{\otimes 6}$ and $(x^3)^{\otimes 9}$ LG models in subsections \ref{ssec:x46} and \ref{ssec:x39}.

\subsection{Flux compactifications}\label{ssec:fluxcomp}
We begin by setting the stage and review the basics of $\mathcal{N}=1$ flux compactifications on Type IIB Calabi--Yau orientifolds. We take the standard geometrical point of view to begin with, but caution that in principle corrections to both the superpotential and K\"ahler potential could enter, since the usual protection arguments rely on having a geometrical description. To overcome these difficulties, we present an argument in section \ref{sec:symmetries} that uses duality symmetries---S-duality of the axio-dilaton and the discrete symmetries of the LG model---to constrain such potential corrections and thereby argue for the existence of critical points even when taking these into account.

\paragraph{Flux superpotential.} Let us begin with the superpotential induced by turning on RR and NS-NS fluxes, denoted by $F_3$ and $H_3$ respectively. The quantization condition of these fluxes corresponds to $F_3,H_3 \in H^3(Y,\mathbb{Z})$, which amounts to
\begin{equation}\label{quantization condition}
    \int_{\Gamma} F_3 \in \mathbb{Z}\, ,  \qquad \int_{\Gamma} H_3 \in \mathbb{Z}\, ,
\end{equation}
for any integral three-cycle $\Gamma \in H_3(Y,\mathbb{Z})$. These fluxes are combined into the complex three-form
\begin{equation}
    G_3 = F_3 - \tau H_3\, ,
\end{equation}
where $\tau= C_0+ie^{-\phi}$ is the axio-dilaton. These fluxes generate a spacetime superpotential for the complex structure moduli $t_i$ as \cite{Gukov:1999ya}
\begin{equation}\label{eq:W}
    W = \int G_3 \wedge \Omega(t_i)\, ,
\end{equation}
where $\Omega$ denotes the holomorphic $(3,0)$-form of the Calabi--Yau threefold. 

\paragraph{K\"ahler potential.} The K\"ahler potential for the complex structure moduli and axio-dilaton decomposes at the classical level as
\begin{equation}\label{eq:K}
    K = K_{\rm a-d}(\tau, \bar\tau)+ K_{\rm cs}(t_i,\bar{t}_i)\, ,
\end{equation}
given by\footnote{By demanding the K\"ahler metric $K_{\tau \bar\tau}$ to be positive-definite, we obtain a bound on the range of $V_E$. To be precise, the strongest constraint is placed at the point $\tau=i$, which requires $-0.22140 \leq \chi/V_E \leq 0.44279$. On the other hand, positivity throughout the rest of the fundamental domain requires $\chi/V_E\geq 0$, since otherwise $e^{-K}$ vanishes. In the LG models of interest we have $\chi<0$, so for convenience we choose to flip the sign of $\chi$ and $V_E$ simultaneously. The resulting bounds then amount to $V_E \gtrsim 406.513$ and $V_E \gtrsim 379.142$ respectively for the $(x^4)^{\otimes 6}$ and $(x^3)^{\otimes 9}$ LG models. }
\begin{equation}\label{eq:Kadcs}
    K_{\rm a-d}(\tau,\bar\tau) =- \log i(\bar \tau-\tau) -2 \log[ V_E + \tfrac{1}{2}\chi E_{\frac{3}{2}}(\tau,\bar\tau)]\, ,\quad K_{\rm cs}(t_i,\bar{t}_i) = -\log i \int \Omega \wedge \bar\Omega \, .
\end{equation}
The first term in $K_{\rm a-d}(\tau,\bar\tau)$ denotes the standard K\"ahler potential for the axio-dilaton, while the second term usually depends on the K\"ahler moduli through the Einstein volume $V_E$. In our case the K\"ahler moduli are absent and therefore $V_E$ is expected to be constant, but note that this term still depends on $\tau$ through the Eisenstein series $E_{\frac{3}{2}}(\tau,\bar\tau)$, which is the $\mathcal{N}=1$ modular completion \cite{Grimm:2007xm} (see also \cite{Green:1997tv, Robles-Llana:2006hby}) of the $\alpha'$-correction \cite{Becker:2002nn, Grimm:2004uq} proportional to the Euler characteristic $\chi$. This function is given by
\begin{equation}
\begin{aligned}
    E_{\frac{3}{2}}(\tau,\bar\tau) &= \sum_{(m,n)\neq (0,0)} \frac{\tau_2^{3/2}}{|m+n\tau|^3} \\
    &= 2\zeta(3) \tau_2^{3/2} + \frac{2\pi^2}{3}\tau_2^{-1/2}+8\pi \frac{\tau_2}\sum_{p\neq 0}\sum_{n=1}^\infty \bigg|\frac{p}{n}\bigg|K_1(2\pi \tau_2 |p|n)e^{2\pi i p n \tau_1}\, .
\end{aligned}
\end{equation}
In the weak-coupling limit $\tau \to i \infty$ we thus find that the K\"ahler potential asymptotes to $K_{\rm a-d}(\tau,\bar\tau) \to -4\log \tau_2$, which agrees with the literature on the LG models \cite{Becker:2007dn, Bardzell:2022jfh}. However, since $E_{\frac{3}{2}}(\tau,\bar\tau)$ is modular invariant, we still find that the K\"ahler potential transforms as $K_{\rm a-d}(\tau,\bar\tau) \to K_{\rm a-d}(\tau,\bar\tau) + \log|c+\tau d|^2$ under SL$(2,\mathbb{Z})$, compensating for the behavior of the superpotential \eqref{eq:W} such that the $\mathcal{N}=1$ supergravity theory is invariant.\footnote{We thank Thomas Grimm and Timm Wrase for discussions on this point.}

\paragraph{Scalar potential.} Having discussed these duality aspects of the K\"ahler potential, let us return to the scalar potential, which is given by the $\mathcal{N}=1$ formula
\begin{equation}\label{eq:Vgen}
    V= e^{K} \left( K^{I\bar J} D_IW D_{\bar J}\bar W - 3 |W|^2 \right)\, ,
\end{equation}
where we take the indices $I,J$ runs over the axio-dilaton and complex structure moduli. We denoted the K\"ahler metric by $K_{I \bar J} = \partial_i \partial_{\bar J} K$ and its inverse by upper indices. The F-terms are given by $D_I W = \partial_i W + K_I W$ with $K_I = \partial_i K$. Supersymmetric extrema such as AdS and Minkowski vacua correspond to $D_I W=0$, with $W\neq 0$ or $W=0$ respectively. In this work we consider both these supersymmetric vacua as well as non-supersymmetric critical points with $\partial_i V=0$ but $D_I W \neq 0$.

\paragraph{Tadpole cancellation condition.} The fluxes $F_3,H_3$ induce a charge $Q_{\rm flux}$ for the D3-brane tadpole that must be canceled globally. Together with the charge contribution $Q_{\rm O3}$ coming from the O3-planes and possible spacetime-filling mobile D3-branes $N_{\rm D3}$ they combine into the tadpole cancellation condition
\begin{equation}\label{eq:tadpole}
    Q_{\rm flux} = \int F_3 \wedge H_3 = Q_{\rm O3}- N_{\rm D3}\, .
\end{equation}
These mobile D3-branes produce additional moduli in the 4d $\mathcal{N}=1$ supergravity theory. In principle the scalar potential should therefore be extremized in these directions as well, as these new scalars could give rise to runaway directions. For this reason we chose to focus on critical points where the flux contribution completely cancels the O3-plane tadpole: $Q_{\rm flux}=Q_{\rm O3}$.

\subsection{Duality symmetries}\label{sec:dualitysymmetries}
Duality symmetries play a central role in this work for finding critical points of flux potentials and arguing for their stability. Let us therefore discuss in detail why we expect 4d $\mathcal{N}=1$ orientifold compactifications to be $SL(2,\mathbb{Z})$ invariant in the first place. In 10d Type IIB string theory this $SL(2,\mathbb{Z})$ symmetry \cite{Schwarz:1995dk, Witten:1995im} relates different regimes for the axio-dilaton $\tau$ under duality transformations
\begin{equation}
    \begin{pmatrix}
        a & b \\
        c & d 
    \end{pmatrix} \in \text{SL}(2,\mathbb{Z}): \qquad \tau \mapsto \frac{a\tau + b}{c\tau + d}\, .
\end{equation}
However, in principle dimensional reduction and orientifold projections may break this symmetry, so it is important to see how it descends to the 4d $\mathcal{N}=1$ supergravity theory, especially in the context of non-geometric models. To begin with, let us consider such a non-geometrical setup from a 10d point of view. In that setting we are free to implement any transformation of SL$(2,\mathbb{Z})$ we would like, and subsequently perform the reduction to the 4d $\mathcal{N}=1$ orientifold theory in each frame. From this perspective we have obtained an SL$(2,\mathbb{Z})$-worth of 4d theories that are all related to one another by dualities from the higher-dimensional point of view. This gives us already a heuristic idea that the SL$(2,\mathbb{Z})$ duality group descends to the 4d $\mathcal{N}=1$ supergravity theory; in the following we make this idea more precise.

\paragraph{4d $\mathcal{N}=2$ compactifications.} Let us next consider 4d $\mathcal{N}=2$ Calabi--Yau compactifications of Type IIB string theory. In this case the axio-dilaton $\tau$ becomes part of the so-called fundamental hypermultiplet under dimensional reduction. Invariance of the hypermultiplet sector under SL$(2,\mathbb{Z})$ is intimately linked to the presence of instanton corrections to the moduli space metric, see \cite{Alexandrov:2013yva} for a review. On the one hand, the assumption of SL$(2,\mathbb{Z})$ invariance has been used to deduce the form of D1-D($-1$) instanton corrections \cite{Robles-Llana:2006hby, Alexandrov:2009qq} and NS5-instantons \cite{Alexandrov:2010ca, Alexandrov:2014mfa, Alexandrov:2014rca}. On the other, SL$(2,\mathbb{Z})$ invariance of D3-instanton corrections \cite{Alexandrov:2012bu,Alexandrov:2012au} has been shown to be the case thanks to modular properties of Donaldson-Thomas invariants and theta series. The presence of SL$(2,\mathbb{Z})$ duality is also intimately linked to the finiteness of the volume of the moduli space \cite{Delgado:2024skw, Vandoren:2025izf}, see \cite{Cortes:2023oje} for a recent example in the mathematics literature.  So these provide ample evidence that SL$(2, \mathbb{Z})$ survives compactification on CY 3-fold.  We should note that since the K\"ahler moduli are hypermultiplets they can mix with the axio-dilaton, which is also a hypermultiplet.  Thus in principle the K\"ahler moduli can pick up transformation properties under SL$(2, \mathbb{Z})$.  This is not the case for complex structure moduli which are vector multiplets and cannot mix with the axio-dilaton and so pick up no transformation property under SL$(2, \mathbb{Z})$.

\paragraph{4d $\mathcal{N}=1$ orientifold compactifications.} Next we consider whether the orientifold projection could remove the SL$(2,\mathbb{Z})$ duality symmetry. In certain cases this can indeed happen: consider for example the orientifold of 10d Type IIB string theory to Type I string theory. The orientifold action projects out the axion of the axio-dilaton, such that only the real dilaton remains. There is therefore no analogue of the $\tau\to\tau+1$ shift symmetry that can be considered, and also S-duality is known to be broken, as Type I string theory at strong-coupling is dual to the heterotic SO(32) string. However, in our case of the 4d $\mathcal{N}=1$ supergravity theory we know that the orientifold projection leaves the axio-dilaton invariant, so for that reason we expect the SL$(2,\mathbb{Z})$ symmetry also to survive. This is furthermore supported by the modular invariance of the expected corrections to the K\"ahler potential, see \cite{Grimm:2007xm} and \eqref{eq:Kadcs}.  Since complex structure moduli are neutral before orientifolding under the SL$(2,\mathbb{Z})$, orientifolding cannot change their transformation properties and so the complex structure moduli continue to be neutral under SL$(2,\mathbb{Z})$ transformations even after the orientifold projection.

\paragraph{Fluxes.} The next question is whether the presence of fluxes can break the SL$(2,\mathbb{Z})$ symmetry. It is well-known that SL$(2,\mathbb{Z})$ acts on the R-R and NS-NS fluxes as
\begin{equation}\label{eq:SL2ZF3H3}
    \begin{pmatrix}
        F_3\\
        H_3
    \end{pmatrix} \mapsto \begin{pmatrix}
        a & b \\
        c & d
    \end{pmatrix}     \begin{pmatrix}
        F_3\\
        H_3
    \end{pmatrix}\, .
\end{equation}
We need to check whether this duality action, which has been checked extensively in weak and strong coupling limits of type IIB survives in the orientifold theory where we have turned on fluxes.
On the flux superpotential \eqref{eq:W} the SL$(2,\mathbb{Z})$ transformation acts as
\begin{equation}\label{eq:SL(2,Z)W}
    W \mapsto \frac{W}{c\tau+d}\, .
\end{equation}
It states that the flux superpotential transforms as a modular form of weight $-1$. In comparison to the transformation of the K\"ahler potential $K \to K + \log|c\tau+d|^2$, we notice that it leaves the $\mathcal{N}=1$ K\"ahler invariant function
\begin{equation}
    \mathcal{G} = K + \log |W|^2\, ,
\end{equation}
invariant. As one can write the scalar potential just in terms of this function $\mathcal{G}$ and its derivatives, we thus find evidence that the 4d $\mathcal{N}=1$ supergravity theory respects the SL$(2,\mathbb{Z})$ even in the presence of fluxes.  Even though there could be corrections to the precise $\tau,\bar \tau$ dependence of $\mathcal{G}$, the fact that fluxes appear up to linear terms in the superpotential is guaranteed as an exact statement: as we go across a BPS domain wall associated to a flux, the superpotential jumps by an amount $\Delta W$ proportional to the tension of the domain wall, which is linear in the fluxes.  So even if there are corrections to $W$ which depend on $\tau$, they cannot destroy this linearity structure of the fluxes. They could in principle introduce flux independent contributions to $W$ and we will allow for this possibility as possible corrections to $W$ from the expected classical form.  Additionally $W$ can in principle receive corrections involving suitable modular objects $f,g,h$, such as 
\begin{equation}\label{eq:Wgen}
    W=\int (f(\tau,t_i) F_3-g(\tau,t_i)\tau H_3)\wedge \Omega(t^i)+h(\tau,t_i)\, ,
\end{equation}
compatible with modular invariance where $t_i$ are the massless chiral fields. Indeed in the geometric examples there are flux independent terms in the superpotential which disappear in the limit of large volume.  However, since in the non-geometric case, one in some sense has a frozen K\"ahler moduli of $O(1)$, then it is natural to expect such terms will be there. We will nevertheless use symmetries to draw conclusions for the existence of vacua even when allowing for such possibilities.

\paragraph{F-theory perspective.} For geometric $\mathcal{N}=1$ orientifold backgrounds, we can also consider the F-theory perspective.  This would correspond to F-theory compactification on a Calabi-Yau fourfold given by $Y_4=(Y_3 \times T^2)/\mathbb{Z}_2$, where $\mathbb{Z}_2$ denotes an involution that acts on the Calabi-Yau threefold $Y_3$ and $T^2$ simultaneously. For this product manifold the complex structure moduli space factorizes into the complex structure moduli space of $Y_3$ (parametrized by $t_{\mathbf{k}}$) and that of $T^2$, where the latter parametrizes the 4d axio-dilaton $\tau$.   To see that the modular symmetry is still a symmetry, we would like to argue that the superpotential terms that would be generated respect it.  To show this it suffices to consider these corrections at the fixed points of the modular group action, as these generate the group, and argue these symmetries are respected by superpotential correction.   Therefore, consider moving to one of the elliptic points $\tau=i, \omega={\rm exp}(2\pi i/3)$ for the complex structure of $T^2$. This background will then respectively have a $\mathbb{Z}_4$ or $\mathbb{Z}_3$ automorphism, which acts on the Calabi-Yau 4-fold by acting as isometry (acting only on the $T^2$ fiber and trivially on $Y_3$). For the flux superpotential as we have already discussed  this corresponds to an exact symmetry. In fact, this symmetry even extends to non-perturbative superpotentials generated by M5-brane instantons \cite{Witten:1996bn} as we will now argue. In the M/F-theory duality picture these instantons come from wrapping M5-branes on divisors of the Calabi-Yau fourfold. Since isometries of the CY 4-fold act on the moduli space of M5-brane instantons, mapping one instanton configuration to another, this symmetry continues to persist even taking these instantons into account.  Even though this argument strictly works in the geometric case, it seems natural to assume to continues in the non-geometric case as well.  For another argument which does not rely on geometric description we consider the finiteness requirement next.

\paragraph{Finiteness.} It was recently argued in \cite{Delgado:2024skw} that finiteness of quantum gravity amplitudes implies that the volume of moduli spaces grows no faster than Euclidean space. For hyperbolic spaces, such as the upper-half plane parametrized by the axio-dilaton, this means that the duality group SL$(2,\mathbb{Z})$ could at most be broken to a finite index subgroup by the orientifold compactification. In particular, transformations such as S-duality---giving rise to the discrete symmetries \eqref{eq:Z4Z3W}---are crucial to ensure that the moduli space volume does not grow exponentially. In contrast to some of the arguments made above, this argument does not rely on the use of a geometric description, as it applies to any moduli space found in a theory of quantum gravity, so also non-geometric constructions and gives further support for the SL$(2,\mathbb{Z})$ invariance.

\subsection{Non-geometric LG models}\label{ssec:nongeom}
For ${\cal N}=1$ supergravity model, it is not an unrealistic expectation to be able to compute the superpotential $W$ exactly as it is a holomorphic section of chiral fields.  
In the context of flux compactifications on CY orientifolds one will naturally get a mixture of K\"ahler and complex structure  moduli in the superpotential as well as in K\"ahler potential since with the reduced ${\cal{N}}=1$ all the chiral fields in principle mix.  This often introduces difficulties in computing the superpotential.  In particular in type IIB setup the complex structure moduli superpotential is far easier to study than the corrections involving K\"ahler moduli.  It was precisely for this reason that in \cite{Becker:2006ks} certain type IIB models were considered which had no K\"ahler moduli.  This naively sounds impossible in geometric models as all CY manifolds have K\"ahler moduli, but mirror symmetry allows for this, as the mirror of IIA on rigid CY manifolds would be precisely of this type, namely IIB with only complex structure moduli, and leads to what one may call non-geometric backgrounds.  Indeed examples of these mirrors were already pointed out in \cite{Vafa:1991uz} where they have a simple description in terms of (2,2) LG models (see also \cite{Candelas:1993nd}).  We will be considering orientifolds of these models together with fluxes turned on them as in \cite{Becker:2006ks}.  We concentrate on two LG models of this type:  One with 6 copies of $x^4$ and the other involving 9 copies of the $x^3$ model.  Both of these models turn out to be special cases of Gepner models \cite{Gepner:1987qi}.  We now turn to studying these LG orientifolds.

\subsection{The \texorpdfstring{$(x^4)^{\otimes 6}$}{x46} LG orientifold model}\label{ssec:x46}
In this section we review the LG model given by six copies of $x^4$, which we will denote by the $(x^4)^{\otimes 6}$ LG model. It corresponds to the $2^6$ Gepner model. its study as a non-geometric background for 4d $\mathcal{N}=1$ orientifolds was initiated in \cite{Becker:2006ks}. In the following we emphasize the symmetries of this theory and their action on the fields and the cycles, since this information will be crucial in the study of critical points of flux potentials in sections \ref{sec:symmetries} and \ref{sec:criticalpoints}. For the reader mainly interested in the 4d $\mathcal{N}=1$ supergravity aspects of these non-geometric backgrounds, we have included a brief summary at the end of this subsection.

\paragraph{The $(x^4)^{\otimes 6}$ LG model.} Let us begin by describing the $(x^4)^{\otimes 6}$ LG model. It has nine scalar fields: six scalars $x_1,\ldots,x_6$ that correspond to the six copies of the $x^4$ model with central charge $c=\tfrac{3}{2}$, and we have chosen to include three additional scalars $z_1,z_2,z_3$ corresponding to trivial minimal models $z^2$ with $c=0$.\footnote{The combination of two $x^4$ models and a $z^2$ model precisely corresponds to the LG model of a two-torus. Also, while all $z^2$ models could be integrated out, keeping at least one saves us from including a $(-1)^F$ in the orbifold action \eqref{eq:orbifold}.} The worldsheet superpotential of this theory is given by
\begin{equation}
    \mathcal{W}=\sum_{i=1}^{6}x_i^4+z_1^2+z_2^2+z_3^2\, .
\end{equation}
While the LG model of interest must be divided by an orbifold action, let us first describe all symmetries of this superpotential. It admits a discrete $\mathbb{Z}_4^6\times \mathbb{Z}_2\rtimes S_6$ symmetry, corresponding to rotating the $x_i$ by a phase $i$, flipping all $z_i$ by a sign,\footnote{Note that the $z_i$ models are all trivial, and hence only the over-all sign of the three models matters.} or permuting the $x_i^4$ models. To be explicit, the action of the $i$-th generator $g_i \in \mathbb{Z}_4^6$ is given by
\begin{equation}\label{eq:Z4^6}
    g_i:\qquad x_j\mapsto i^{\delta_{ij}}x_j\, ,\quad z_i\mapsto z_i\, ,
\end{equation}
while the $\mathbb{Z}_2$ acts as
\begin{equation}\label{eq:Z2}
    g_0:\qquad x_j\mapsto x_j\, ,\quad z_i\mapsto -z_i\, .
\end{equation}
The $(x^4)^{\otimes 6}$ LG model is defined as the 2d $\mathcal{N}=(2,2)$ superconformal field theory (SCFT) divided by the orbifold action $g=g_0 \cdots g_6 $, which is a combination of the above symmetries. It acts on the coordinates as
\begin{equation}\label{eq:orbifold}
   g : \qquad x_i\mapsto ix_i\, ,\quad z_i\mapsto-z_i\, .
\end{equation}
The orientifold action $\sigma$ corresponds to a symmetry  which sends $\mathcal{W}(\sigma x, \sigma z) = - \mathcal{W}(x,\sigma)$ \cite{Brunner:2003zm,Hori:2006ic}. The possible orientifold actions for the $(x^4)^{\otimes 6}$ were classified in \cite{Becker:2006ks}, out of which we choose the canonical choice
\begin{equation}\label{eq:orientifold}
    \sigma:\qquad x_i\mapsto e^{\frac{2\pi i}{8}}x_i\, ,\quad z_i\mapsto iz_i\, ,
\end{equation}
{\color{red}dressing with the quantum symmetry} as was also studied recently in \cite{Rajaguru:2024emw, Becker:2024ayh}. This orientifold action squares to the orbifold symmetry \eqref{eq:orbifold} of the LG model, i.e.~$\sigma^2=g$. It induces the largest O3-plane tadpole contribution out of the possible choices {\color{red}\cite{Morros:2026yyi}, given by\footnote{\color{red} Compared to the original paper \cite{Becker:2006ks} which gave $Q_{\rm O3}=40$, the recent paper \cite{Morros:2026yyi} pointed out a mistake in their normalization of the point class, lowering the O3-plane charge by a factor of two.}
\begin{equation}
    Q_{\rm O3} = 20\, .
\end{equation}}
\paragraph{Cohomology ring.} Under the LG/CY correspondence, deformations of the compact space---corresponding to moduli of the 4d supergravity theory---can be described by deformations of the worldsheet superpotential. To be precise, the cohomology ring is given by the chiral-chiral ring of the 2d $\mathcal{N}=(2,2)$ SCFT
\begin{equation}
    \mathcal{R} = \bigg[\frac{\mathbb{C}[x_1,...,x_6]}{\sum_{i=1}^{6}(\partial_{x_i}\mathcal{W})}\bigg]^{\mathbb{Z}_4}\, .
\end{equation}
This is a 182-dimensional complex vector space spanned by the monomials
\begin{equation}\label{eq:monomials}
    x^{\mathbf{k}}=x_1^{k_1}\cdots x_{6}^{k_6}\, ,
\end{equation}
with coefficients given by
\begin{equation}\label{eq:ks}
    \mathbf{k}=(k_1,\ldots, k_6)\, : \qquad k_i\in\{0,1,2\}\, , \quad \sum k_i\equiv 0\mod 4\, ,
\end{equation}
where the last condition ensures that we consider only monomials that are invariant under the orbifold action \eqref{eq:orbifold}. The exponents $k_i$ of the monomials $x^{\mathbf{k}}$ correspond to the U(1)-charges of the states. In terms of the cohomology ring, the charge decomposition $\sum k_i=0,4,8,12$ corresponds to the Hodge decomposition into the spaces $H^{3,0},H^{2,1},H^{1,2},H^{0,3}$ of $(p,q)$-forms. The complex structure deformations are controlled by the $(2,1)$-forms, and hence correspond to the 90 monomials $x^{\mathbf{k}}$ with $\sum k_i=4$. At the level of the 2d $\mathcal{N}=(2,2)$ SCFT they correspond to the marginal deformations. Denoting these deformation parameters by $t_{\mathbf{k}}$, we can write the deformed worldsheet superpotential as
\begin{equation}\label{eq:Wdeform}
    \mathcal{W}(x_i;t_{\mathbf{k}})=\sum_{i=1}^{6}x_i^4-\sum_{\mathbf{k}}t_{\mathbf{k}}x^{\mathbf{k}}+z_1^2+z_2^2+z_3^2\, .
\end{equation}
where the sum is restricted to $\mathbf{k}$ with $\sum k_i = 4$. Let us next consider how the symmetries \eqref{eq:Z4^6} of the LG model act on these deformation moduli. Requiring the worldsheet superpotential $\mathcal{W}$ to remain invariant, we find that the moduli transform as
\begin{equation}\label{eq:gtx46}
    g_i:\qquad t_{\mathbf{k}}\mapsto i^{-k_i}t_{\mathbf{k}}\, .
\end{equation}
In order to determine the even cohomologies in the LG orbifold, we need to consider the twisted sectors under $e^{-2\pi i J_0}$ \cite{Vafa:1989xc}: here only the states corresponding the $(0,0)$ and $(3,3)$-form survive, and all others are projected out. Therefore, the Hodge diamond of the LG orbifold is given by:\\
\begin{equation}\label{eq:hodgediamond}
\dim(H^{p,q}) = \begin{tikzpicture}[baseline={([yshift=-.5ex]current bounding box.center)},scale=0.7,cm={cos(45),sin(45),-sin(45),cos(45),(15,0)}]
\draw (0,0) node {$1$};
\draw (1,0) node {$0$};
\draw (2,0) node {$0$};
\draw (3,0) node {$1$};

\draw (0,1) node {$0$};
\draw (1,1) node {$0$};
\draw (2,1) node {$90$};
\draw (3,1) node {$0$};

\draw (0,2) node {0};
\draw (1,2) node {$90$};
\draw (2,2) node {$0$};
\draw (3,2) node {$0$};

\draw (0,3) node {$1$};
\draw (1,3) node {$0$};
\draw (2,3) node {$0$};
\draw (3,3) node {$1$};
\end{tikzpicture}\,  .
\end{equation}
The vanishing of $h^{11}=0$ is what makes this model non-geometrical, as it has no K\"ahler moduli to parametrize the volume of the space. We also note that all $(p,q)$-forms are odd under $\sigma$ and therefore survive the orientifold projection \eqref{eq:orientifold}. In conclusion, the resulting 4d $\mathcal{N}=1$ supergravity theory has 91 chiral fields, corresponding to $90$ complex structure moduli $t_{\mathbf{k}}$ and the axio-dilaton $\tau$.

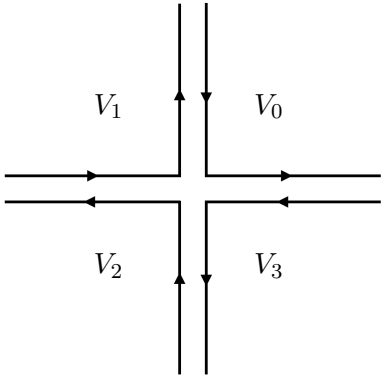
\begin{figure}
    \centering
    \includegraphics[width=0.35\linewidth]{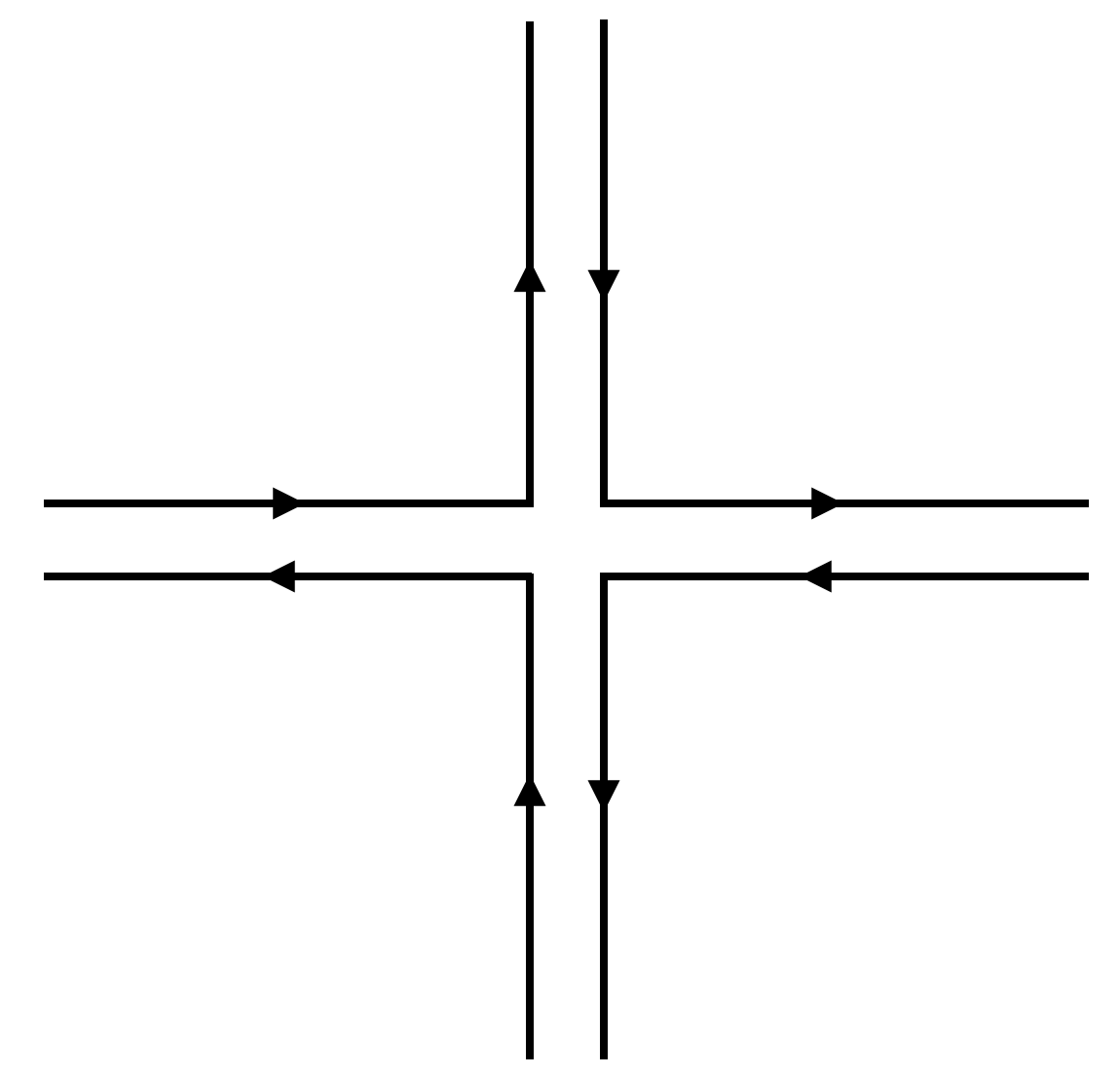}
    \begin{picture}(0,0)\vspace*{-1.2cm}
    \put(-55,100){$V_0$}
    \put(-55,40){$V_3$}
    \put(-115,40){$V_2$}
    \put(-115,100){$V_1$}
    \end{picture}\vspace*{-0.1cm}
    \caption{The four contours $V_0,V_1,V_2,V_3$ of the $\mathcal{W}=x^4$ LG model in the complex $x$-plane.}
    \label{fig:contours}
\end{figure}

\paragraph{Homology of the $x^4$ LG model.} Given the field content of the model, let us next construct the basis for the fluxes $F_3,H_3$. Since these fluxes are quantized \eqref{quantization condition}, we need to find an integral three-cycle homology basis for the LG model. The corresponding objects in the LG model are given by the A-type D-branes \cite{Hori:2000ck}. Let us begin with a single copy $\mathcal{W}=x^4$ in the LG model. The A-type D-branes are given by contours in the $x$-plane where the phase of the worldsheet superpotential $\mathcal{W}$ is constant. By an R-rotation, we can describe this as the pre-image of the positive real line as
\begin{equation}
    \mathrm{Im}(\mathcal{W})=0\, .
\end{equation}
As illustrated in figure \ref{fig:contours}, we choose as generators the four wedges $V_0,V_1,V_2,V_3$ of the $x$-plane.\footnote{Strictly speaking, we should tensor these cycles $V_0,V_1,V_2,V_3$ with the A-type D-branes of the trivial $z^2$ LG model. However, since these wedges would be the upper-half plane and lower-half plane in the $z$-plane, and these would be linearly dependent in a way similar to \eqref{eq:lindep}, it would just correspond to tensoring with a one-dimensional vector space and therefore can be left out. \label{footnote:z2wedges}} These wedges are linearly dependent as described by the relation
\begin{equation}\label{eq:lindep}
    V_0+V_1+V_2+V_3=0\, .
\end{equation}
The intersection $\langle V_m | V_n \rangle$ between the wedges has been computed geometrically \cite{Hori:2000ck}
\begin{equation}
    \langle V_{m}|V_{n}\rangle=\delta_{m,n}-\delta_{m+1,n}\, ,
\end{equation}
with $m,n\in \mathbb{Z}_4$. The $\mathbb{Z}_4$ symmetry acting on the LG coordinate $g: x \mapsto i x$ acts by permuting these wedges as
\begin{equation}\label{eq:Z4V}
    g \in \mathbb{Z}_4:\qquad V_{n}\mapsto V_{n+1}\, ,
\end{equation}
where again $n \in \mathbb{Z}_4$. Given this integral homology basis, let us next turn to the intersection with the cohomology basis. For the single copy $x^4$ LG model, the cohomology basis is described by the chiral-chiral ring $\mathcal{R} = \mathbb{C}[x]/x^3$, which is spanned by the monomials $1,x,x^2$. These monomials $x^k$ correspond to the R-R ground states $|\ell \rangle$ according to $\ell=k+1=1,2,3$. The integral of such a one-form over a wedge is then given by the contour integral \cite{Hori:2000ck}
\begin{equation}\label{eq:overlap}
    \langle V_n|\ell \rangle=\int_{V_n} x^{\ell-1}e^{-\mathcal{W}}dx=\frac{1}{4}i^{n\ell }(1-i^{\ell})\Gamma\left(\frac{\ell}{4}\right)\, .
\end{equation}
with $k=0,1,2$. To simplify the expansion of the superpotential, it is convenient to renormalize the cohomology basis as
\begin{equation}
    |\chi_\ell \rangle  = \frac{4}{\Gamma\left(\frac{\ell}{4}\right)} | \ell \rangle\, .
\end{equation}
The symmetries of the fluxes are most conveniently analyzed in terms of this basis. This becomes apparent by writing these states in terms of the wedges as
\begin{equation}\label{eq:chitoV}
    |\chi_\ell \rangle = \sum_n i^{n \ell} |V_n \rangle\, .
\end{equation}
Since the $\mathbb{Z}_4$ symmetry permutes the wedges according to \eqref{eq:Z4V}, this expansion tells us that the R-R ground states are eigenstates of the symmetry $g \in \mathbb{Z}_4$, namely
\begin{equation}
   g : \quad |\chi_\ell \rangle \to i^\ell |\chi_\ell \rangle\, .
\end{equation}
The overlap between two R-R ground states $\langle \chi_{\ell} |\chi_{\ell'} \rangle$ can be computed from \eqref{eq:overlap} by passing to the homology basis through \eqref{eq:chitoV} as
\begin{equation}\label{intersection}
            \langle \chi_{\ell'}|\chi_{\ell} \rangle =4\delta_{\ell'+\ell,4}(1-i^{\ell})\, .
\end{equation}
Complex conjugation relates these R-R ground states as
\begin{equation}
    \overline{\chi_\ell} = \chi_{4-\ell}\, ,
\end{equation}
as follows again by using the expansion \eqref{eq:chitoV} into wedges $V_n$ and using their reality.

\paragraph{Homology of the $(x^4)^{\otimes 6}$ LG model.} With the knowledge of the single $x^4$ model, we can now tensor $6$ copies of them and the trivial $z_i^2$ models to get the complete story. We obtain the R-R ground states by taking tensor products of the ground states of the $x^4$ LG model
\begin{equation}\label{eq:tensor}
    |\chi_{\bell} \rangle = |\chi_{\ell_1}\rangle \otimes |\chi_{\ell_2}\rangle \otimes...\otimes | \chi_{\ell_6} \rangle 
\end{equation}
where $\bell=(l_1,...,l_6)$ with $\ell_i\in\{1,2,3\}$. Let us next describe how the discrete symmetry $\mathbb{Z}_4^6 \times \mathbb{Z}_2$ acts on these states, both for the purpose of characterizing the symmetries and performing the orbifold projection. The phase rotation by the $i$-th generator $g_i \in \mathbb{Z}_4^6$ acts as
\begin{equation}\label{eq:g_ichi}
    g_i:\chi_{\bell}\mapsto i^{-\ell_i}\chi_{\bell}  \, .
\end{equation}
The operator $g_0 \in \mathbb{Z}_2$ that reflects the coordinates $z_i$ of the trivial LG models acts by\footnote{This can be seen from the fact that technically speaking the $z_i^2$ LG models contribute a tensor factor to \eqref{eq:tensor} corresponding to a state in a one-dimensional space. This state then picks up a minus sign under the reflection $z_i \mapsto -z_i$. See also footnote \ref{footnote:z2wedges} for a description of the homology of the $z^2$ LG model.}
\begin{equation}
    g_0:\chi_{\bell}\mapsto -\chi_{\bell}\, .
\end{equation}
Invariant states under the orbifold projection $g=g_0 g_1 \cdots g_6$ then correspond to RR charges that add up to
\begin{equation}\label{orbifold projection l}
    \sum_{i=1}\ell_i\equiv2 \mod 4\, ,
\end{equation}
which is analogue of the condition \eqref{eq:ks} on the monomial deformations $x^{\mathbf{k}}$ in the chiral-chiral ring. Similar to these deformations, there is a correspondence between the ground states with $\sum_{i=1}^{6}\ell_i=6,10,14,18$ and the Hodge decomposition into $(p,q)$-form spaces $H^{3,0},H^{2,1},H^{1,2},H^{0,3}$, respectively; all of these correspondences are summarized in table \ref{table:LGmodel}. 
\begin{table}[]
    \centering
    \begin{tabular}{| c || c | c | c | c |}\hline
        $H^{p,q}$ & $H^{3,0}$ & $H^{2,1}$ & $H^{1,2}$ & $H^{0,3}$\\ \hline 
        $\sum_i \ell_i$ & 6 & 10 & 14 & 18 \\ \hline
         & & $(1,1,2,2,2,2)$ & $(3,3,2,2,2,2)$ & \\
        $\bell$ & $(1,1,1,1,1,1)$ &   $(1,1,1,2,2,3)$ &   $(3,3,3,2,2,1)$  & $(3,3,3,3,3,3)$ \\
         & & $(1,1,1,1,3,3)$ & $(3,3,3,3,1,1)$ & \\ \hline
    \end{tabular}
    \caption{Summary of the Hodge decomposition into $(p,q)$-form spaces of the $(x^4)^{\otimes 6}$ LG model. We have listed $S_6$ representatives of the RR charges of the ground states \eqref{eq:tensor} spanning these spaces.}
    \label{table:LGmodel}
\end{table}

Similar to the cohomology, the homology is generated by the tensor product of wedges $V_n$ with equivalence relations
\begin{equation}\label{eq:gamma}
    \gamma_{\mathbf{n}} \sim [V_{\mathbf{n}}] \sim [-V_{\mathbf{n+1}}] \sim [V_{\mathbf{n+2}}]\sim[-V_{\mathbf{n+3}}]\, ,
\end{equation}
where $\mathbf{n}=(n_1,...,n_6)$ are six integers $n_i\in \{0,1,2,3\}$ with
\begin{equation}
    |V_{\mathbf{n}}\rangle=|V_{n_1}\rangle\otimes|V_{n_2}\rangle\otimes|V_{n_3}\rangle\otimes|V_{n_4}\rangle\otimes|V_{n_5}\rangle\otimes|V_{n_6}\rangle \, .
\end{equation}
While this basis is overcomplete, following \cite{Becker:2024ayh} a complete basis is obtained by restricting $n_i \in \{0,1,2\}$ considering the first 182 integers in base 3, i.e.~the range $\mathbf{n}=(0,0,0,0,0,0)$ to $\mathbf{n}=(0,2,0,2,0,1)$. 

Using these bases $|\chi_{\bell}\rangle$ for the cohomology and $\langle \gamma_{\mathbf{n}}|$ for the integral homology lattice, let us next turn to the products between these states, as we need these for evaluating the tadpole and the spacetime superpotential. First, the intersection between the orbifold homology can be computed by
\begin{equation}\label{eq:orbifold intersection}
    \langle \gamma_{\mathbf{n'}}|\gamma_{\mathbf{n}}\rangle=\langle V_{\mathbf{n'}}|V_{\mathbf{n}}\rangle-\langle V_{\mathbf{n'+1}}|V_{\mathbf{n}}\rangle+\langle V_{\mathbf{n'+2}}|V_{\mathbf{n}}\rangle-\langle V_{\mathbf{n'+3}}|V_{\mathbf{n}}\rangle\,.
\end{equation}
Here the left hand side is the intersection form of the orbifold LG model and the right hand side is the intersection before orbifolding, i.e.~just tensor product of $6$ copies of a single $x^4$ model.\footnote{In the literature various conventions have been used for the homology basis $\gamma_{\mathbf{n}}$, like the invariant sum in \cite{Rajaguru:2024emw} or with a normalization factor $4^{-1/2}$ in \cite{Becker:2006ks}. All these procedures result in the same intersection pairing \eqref{eq:orbifold intersection} for the orbifold LG model in terms of the non-orbifold basis $V_{\mathbf{n}}$.} Using this relation, the intersection between the RR ground states and the integral homology basis is given by
\begin{equation}\label{eq:quant}
        \langle \gamma_\mathbf{n}|\chi_{\bell}\rangle =i^{\mathbf{n}\cdot\bell}\prod_{i=1}^{6}(1-i^{\ell_i})    \, ,
\end{equation}
and the intersection pairing for the RR ground states reads
\begin{equation}
        \langle\chi_{\bell'}|\chi_{\bell}\rangle =4^5\delta_{\bell+\bell',4}\prod_{i=1}^{6}(1-i^{\ell _i})\, .
\end{equation}
Finally, we want to see how the discrete symmetry of the LG model acts on the three-form flux $H$ (which can be $F_3$ or $H_3$), so that we can use those symmetry to describe the selection rule. Let us first expand the three-form flux into the R-R ground states as
\begin{equation}\label{eq:Hexpand}
    H=\sum_{\bell}H^{\bell}\chi_{\bell}\, .
\end{equation}
The symmetry action \eqref{eq:g_ichi} of $g_i \in \mathbb{Z}_4$ on the R-R ground states induces an action on the flux components $H^\bell$ given by
\begin{equation}
    g_i:H^{\bell}\mapsto i^{-\ell_i}H^{\bell}\, .
\end{equation}

\paragraph{Flux superpotential.} Let us next consider the classical spacetime superpotential that is induced by the fluxes. For simplicity, let us turn on only a single flux component $G_3 = \gamma_{\mathbf{n}}$. The superpotential at the LG point in the moduli space is then computed by the overlap $W = \langle \gamma_{\mathbf{n}} | \mathbf{1} \rangle$, since $|\mathbf{1}\rangle $ corresponds geometrically to the $(3,0)$-form, cf.~table \ref{table:LGmodel}. However, we are interested in the moduli dependence of the superpotential, so we want to consider the case where the worldsheet superpotential is deformed \eqref{eq:Wdeform}. This yields as moduli-dependent superpotential
\begin{equation}\label{eq:Wfull}
    W(\gamma_{\mathbf{n}}, t_{\mathbf{k}})= \int_{\gamma_{\mathbf{n}}} d^6 x\ e^{-\sum_{i=1}^6 x_i^4 + \sum_{\mathbf{k}}x^{\mathbf{k}}t_{\mathbf{k}}}\, .
\end{equation}
This integral describes the classical flux superpotential to all orders in the complex structure moduli $t_{\mathbf{k}}$, which may be obtained as in \eqref{eq:Wall} by expanding the integrand. 

For now, let us focus on the symmetry transformations of this flux superpotential. Under the $i$-th generator $g_i \in \mathbb{Z}_4^6$ we find that the superpotential only picks up a factor of $i$ under the transformation of the measure $d^6 x$, as the deformed worldsheet superpotential $\mathcal{W}(x_i, t_{\mathbf{k}})$ in the exponent is invariant, provided we also relabel the cycle as $\gamma_{\mathbf{n}} \mapsto \gamma_{\mathbf{n}+\mathbf{1}}$. Since this is just how the integral basis transforms under $\mathbb{Z}_4^6$, we may write it in general as
\begin{equation}
    g_i \in \mathbb{Z}_4^6 : \quad W(G_3, t_{\mathbf{k}})\mapsto W(g_i \cdot G_3, i^{-k_i}t_{\mathbf{k}}) = i W(G_3, t_{\mathbf{k}})\, .
\end{equation}
In other words, the spacetime flux superpotential has charges $\mathbf{1}=(1,1,1,1,1,1)$, matching with the RR charges $\mathbf{\ell}=\mathbf{1}$ of the holomorphic $(3,0)$-form.

Given these transformation properties, let us return to the expansion of the superpotential \eqref{eq:Wfull}, for which an all-order expression was derived in \cite{Rajaguru:2024emw}. Considering for convenience now a complex three-form flux along a single component $G_3=\chi_{\mathbf{l}}$, the term at order $t_{\mathbf{k}_1}\cdots t_{\mathbf{k}_r}$ follows from
\begin{equation}\label{eq:Wall}
    \partial_{t_{\mathbf{k}_1}}\cdots \partial_{t_{\mathbf{k}_r}} W(\chi_{\bell},t_{\mathbf{k}}) \big|_{t_{\mathbf{k}=0}} = \frac{1}{2}\delta(\mathbf{L}+\bell \text{ mod } 4) \prod_{i=1}^6 (1-i^{L_i})\Gamma\left(\frac{L_i}{4}\right)\,  ,
\end{equation}
where we defined $\mathbf{L} = \sum_{\alpha=1}^r\mathbf{k}_\alpha + 1$. Notice that this expression may be used to compute any period of the $(3,0)$-form in the complex three-form basis $\chi_{\bell}$, or equivalently in the integral basis $\gamma_{\mathbf{n}}$ by using \eqref{eq:quant}. This means that \eqref{eq:Wall} does not only enable us to compute the classical flux superpotential to all orders, but also the classical K\"ahler potential of the 4d $\mathcal{N}=1$ supergravity theory. We can therefore use these expressions to compute the scalar potential in this classical approximation. Although the conclusions we derive from these results should be taken lightly, it allows us to give classical values for quantities such as the cosmological constant, moduli masses, and numbers of tachyons in section \ref{sec:criticalpoints}.

\paragraph{Axio-dilaton and $\mathrm{SL}(2,\mathbb{Z})$ symmetry.} Most of the discussion here focused on the complex structure moduli $t_{\mathbf{k}}$ and the corresponding discrete symmetry $\mathbb{Z}_4^6$ (or $\mathbb{Z}_3^9$ for the $(x^3)^{\otimes 9}$ LG model), but we did not yet discuss the axio-dilaton $\tau$ or the accompanying SL$(2,\mathbb{Z})$ symmetry we established in section \ref{sec:dualitysymmetries}. In principle we are free to fix $\tau$ to either of the elliptic points $\tau=i,\exp(2\pi i/3)$ and use the accompanying discrete symmetry $\mathbb{Z}_3,\mathbb{Z}_4 \subset \text{SL}(2,\mathbb{Z})$. However, as was readily shown in \cite{Becker:2006ks}, for supersymmetric flux vacua the quantization condition on the fluxes requires us to set $\tau=i$ in the $(x^4)^{\otimes 6}$ LG model, and similarly $\tau=\exp(2\pi i/3)$ in the $(x^3)^{\otimes 9}$ LG model. For this reason we make the natural choice of elliptic point for the axio-dilaton $\tau$ in both of the LG models.

\paragraph{Summary of supergravity aspects.} To conclude this review, let us summarize the main features of the $(x^4)^{\otimes 6}$ LG model in light of the 4d $\mathcal{N}=1$ supergravity theory:
\begin{itemize}
    \item The theory has $91$ chiral fields in total: one axio-dilaton $\tau$ and $90$ complex structure moduli $t_{\mathbf{k}}$. The complex structure moduli come from the monomial deformations \eqref{eq:monomials} of the worldsheet superpotential, with their exponents $\mathbf{k}=(k_1,\ldots,k_6)$ valued in $k_i\in\{0,1,2\}$ and constrained to $\sum_i k_i=4$ by the orbifold projection. 
    \item The orientifold projection \eqref{eq:orientifold} induces an O3-plane charge given by {\color{red}$Q_{\rm O3} = 20$}. We restrict our attention in this work to fluxes that saturate the corresponding tadpole bound, to avoid any moduli coming from mobile D3-branes. Also, all three-forms are odd under the orientifold involution, so none are projected out, and similarly neither are the complex structure moduli nor the axio-dilaton.
    \item We describe the integral three-form basis for the fluxes in terms of a complete basis $\gamma_{\mathbf{n}}$ as defined in \eqref{eq:gamma}, with $\mathbf{n}=(n_1,\ldots,n_6)$ and $n_i \in \{0,1,2\}$ ranging from $\mathbf{n}=(0,0,0,0,0,0)$ to $\mathbf{n}=(0,2,0,2,0,1)$ when interpreted as a base-3 integer.
    \item We consider a complex three-form basis $\chi_{\bell}$ defined by \eqref{eq:tensor} with RR charges $\bell=(\ell_1,\ldots,\ell_6)$, which are valued in $\ell_i \in \{1,2,3\}$ and constrained to $\sum_i \ell_i\equiv 2 \mod 4$ by the orbifold projection. The decomposition into eigenspaces according to $\sum_i \ell_i =6,10,14,18$ corresponds precisely to the Hodge decomposition, which has been summarized in table \ref{table:LGmodel}. We expand a three-form flux $H$ as described in \eqref{eq:Hexpand} into components $H^{\bell}$ in this basis. We can relate this complex three-form basis to the integral basis $\gamma_{\mathbf{n}}$ through \eqref{eq:quant}, which in turn allows us to check the quantization of the flux quanta.
    \item The theory possesses a $\mathbb{Z}_4^6 \rtimes S_6$ discrete gauge symmetry that it inherits from the $(x^4)^{\otimes 6}$ LG model. The permutation symmetry $S_6$ acts by permuting the charges of the moduli $t^{\mathbf{k}}$ and flux quanta $H^{\bell}$. The $i$-th symmetry operator $g_i \in \mathbb{Z}_4^6$ acts by a phase multiplication as
    \begin{equation}
        g_i: t_{\mathbf{k}} \mapsto i^{-k_i}t_{\mathbf{k}}\,,\qquad H^\bell\mapsto i^{-\ell_i}H^\bell\, .
    \end{equation}
    We also found that the flux superpotential transforms the same way as the holomorphic $(3,0)$-form does
    \begin{equation}
    g_i \in \mathbb{Z}_4^6 : \quad W(G_3, t_{\mathbf{k}})\mapsto  i W(G_3, t_{\mathbf{k}})\, ,
    \end{equation}        
    corresponding to an eigenstate with charges $\mathbf{1}=(1,1,1,1,1,1)$. These symmetry transformations play a crucial role in this work, as we use them in the next section \ref{sec:nongeometric} to derive the selection rules for the superpotential and scalar potential.
\end{itemize}

\subsection{The \texorpdfstring{$(x^3)^{\otimes 9}$}{x39} LG orientifold model}\label{ssec:x39}
Now, let us describe another LG model, which is also a mirror of a rigid Calabi-Yau compactification: the $(x^3)^{\otimes 9}$ LG model, also known as the $1^9$ Gepner model. Its study as a non-geometric background for 4d $\mathcal{N}=1$ orientifolds was also initiated in \cite{Becker:2006ks}. In the following, we summarize the corresponding ingredients in the $(x^3)^{\otimes 9}$ LG model, i.e.~the moduli, (co)homology and symmetries of the model to establish the selection rules as in the $(x^{4})^{\otimes 6}$ LG model. We adopt different conventions compared to the recent papers \cite{Becker:2023rqi, Becker:2024ijy} in order to match with the $(x^4)^{\otimes 6}$ LG model discussed above.

\paragraph{The $(x^{3})^{\otimes 9}$ LG model.} The $(x^3)^{\otimes 9}$ LG model has nine scalar fields $x_1,...,x_9$. It is a tensor product of nine copies of the $x^3$ model, which is a minimal model with central charge $c=1$. The worldsheet superpotential is then given by
\begin{equation}
    \mathcal{W}=\sum_{i=1}^{9}x_i^3  \,.
\end{equation}
While the LG model of interest must be divided by an orbifold action, we first describe all symmetries of this superpotential: it admits a $\mathbb{Z}_3^9\rtimes S_9$ symmetry, corresponding to rotating the $x_i$ with a phase $\omega =\exp(2\pi i/3)$ and permuting the nine $x_i^3$ models. Explicitly, the action of the $i$-th generator $g_i \in \mathbb{Z}_3^9$ is given by
\begin{equation}\label{eq:Z3^9}
    g_i:x_j\mapsto\omega^{\delta_{ij}}x_j  \,.
\end{equation}
Then, the $(x^3)^{\otimes 9}$ LG model is the 2d $\mathcal{N}=(2,2)$ SCFT obtained by orbifolding by the diagonal symmetry  $g=g_1g_2g_3...g_9$, which acts on the scalar field $x_i$ by:
\begin{equation}
    g:x_i\mapsto \omega x_i  \,.
\end{equation}
As in the case of the $(x^4)^{\otimes 6}$ LG model, we need to orientifold the model to be able to turn on fluxes and break half the supersymmetry. The possible orientifold actions are again classified in \cite{Becker:2006ks}, and we choose the orientifold action
\begin{equation}\label{eq:x3orientifold}
    \sigma: (x_1,x_2,x_3,x_4,x_5,x_6,x_7,x_8,x_9)\mapsto -(x_2,x_1,x_3,x_4,x_5,x_6,x_7,x_8,x_9) \,,
\end{equation}
which is the case also studied recently in \cite{Becker:2023rqi,Becker:2024ijy}. The orientifold action squares to the identity and acts on the worldsheet superpotential as $\mathcal{W}(\sigma x)=-\mathcal{W}(x)$. This choice of orientifold induces the largest O3-plane charge among all of the possibilities described in \cite{Becker:2006ks}, given by
\begin{equation}
    Q_{\rm O3}=12 \,.
\end{equation}

\paragraph{Cohomology ring.} We can determine the cohomology of the $(x^3)^{\otimes 9}$ LG model in the same way as the $(x^4)^{\otimes 6}$ LG model through the standard methodology; the details can be found in \cite{Becker:2006ks}. The middle cohomology is given by the chiral-chiral ring of the 2d $\mathcal{N}=(2,2)$ SCFT as
\begin{equation}
    \mathcal{R}=\Big[\frac{\mathbb{C}[x_1,...,x_9]}{\sum_{i=1}^{9}(\partial_{x_i}\mathcal{W})}\Big]^{\mathbb{Z}_3} \,.
\end{equation}
This is an 170-dimensional complex vector space spanned by the monomials
\begin{equation}
    x^{\mathbf{k}}=x_1^{k_1}x_2^{k_2}...x_{9}^{k_9} \,,
\end{equation}
with exponents given by
\begin{equation}
    \mathbf{k}= (k_1,\ldots, k_9)\ : \qquad k_i\in\{0,1\}\, , \quad \sum_{i=1}^{9}k_i\equiv 0\mod 3\, ,
\end{equation} 
where the last condition ensures that we consider only monomials invariant under the $\mathbb{Z}_3$ orbifolding. The Hodge decomposition relates monomials with $\sum k_i=0,3,6,9$ to the spaces $H^{3,0},H^{2,1},H^{1,2},H^{0,3}$ in the middle cohomology. In particular, the complex structure deformations are controlled by the $(2,1)$-forms, corresponding to the $84$ monomials $x^{\mathbf{k}}$ with $\sum k_i=3$. Under the LG/CY correspondence, these complex structure moduli correspond to deformations of the worldsheet superpotential, given explicitly by
\begin{equation}
    \mathcal{W}(x_{i};t_{\mathbf{k}})=\sum_{i=1}^{9}x_i^3-\sum_{\mathbf{k}}t_{\mathbf{k}}x^{\mathbf{k}} \,,
\end{equation}
where the sum is restricted to $\mathbf{k}$ with $\sum k_i = 3$. Let us next consider how the $\mathbb{Z}_3^9$ symmetry given by \eqref{eq:Z3^9} acts on these deformations. Requiring the worldsheet superpotential to be invariant, we find that the moduli must transform as
\begin{equation}\label{eq:gtx39}
    g_i:t_{\mathbf{k}}\mapsto \omega^{-k_i}t_{\mathbf{k}} \,.
\end{equation}
Similar to the $(x^4)^{\otimes 6}$ LG model, the twisted sectors only contribute the $(0,0)$ and $(3,3)$-form, so only the middle cohomology is non-trivial. The vanishing of $h^{1,1}$ again tells us that this model is non-geometrical as it has no K\"ahler modulus. 

\begin{figure}
    \centering
    \includegraphics[width=0.35\linewidth]{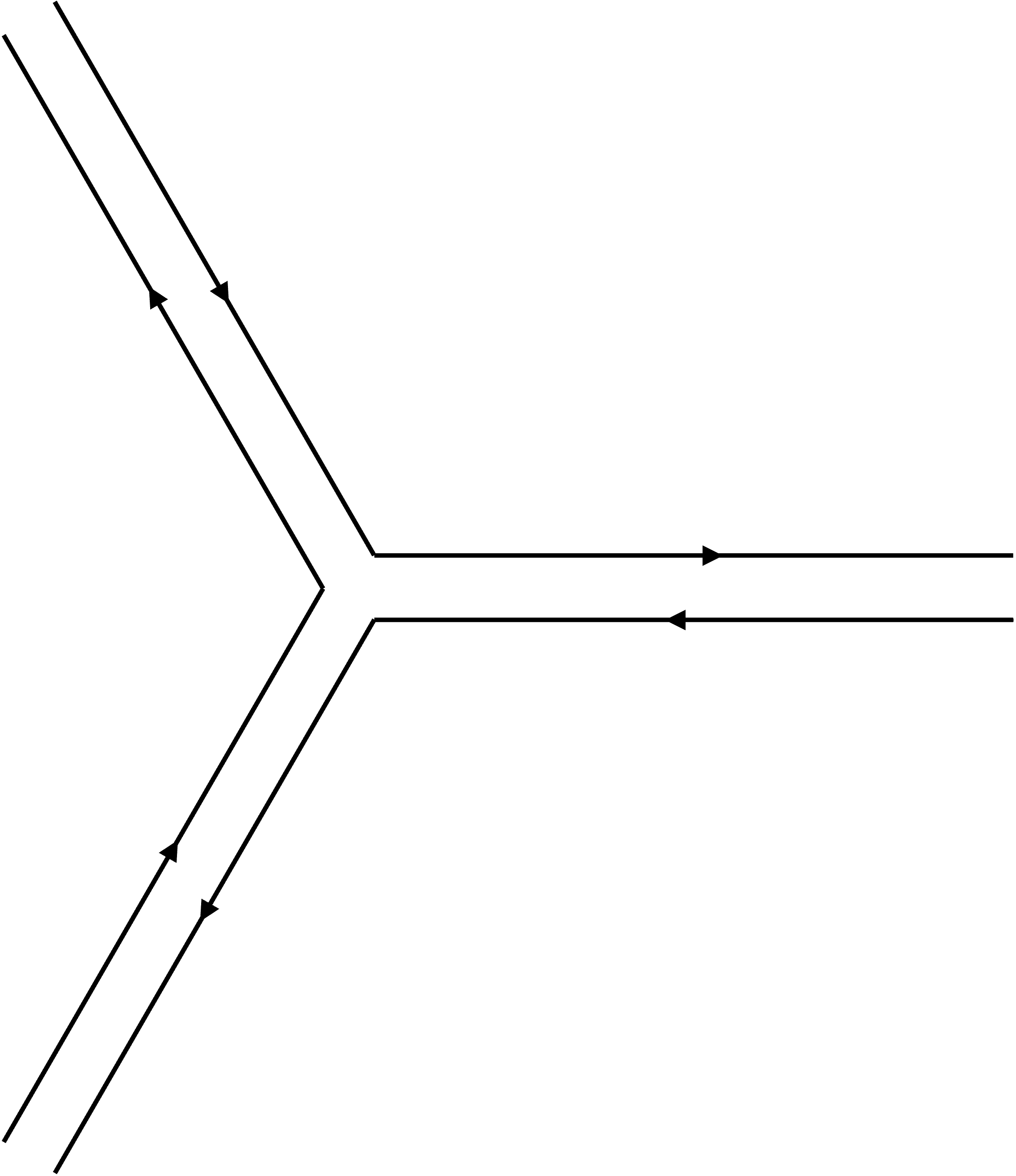}
    \begin{picture}(0,0)\vspace*{-1.2cm}
    \put(-80,120){$V_0$}
    \put(-80,40){$V_2$}
    \put(-160,80){$V_1$}
    \end{picture}\vspace*{-0.1cm}
    \caption{The three contours $V_0,V_1,V_2$ of the $\mathcal{W}=x^3$ LG model in the complex $x$-plane.}
    \label{fig:contours2}
\end{figure}
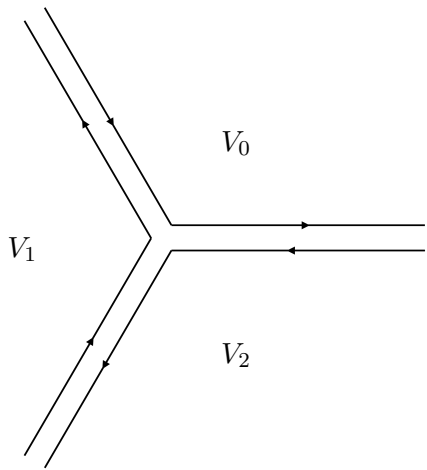

\paragraph{Homology of the $x^3$ LG model.} Given the field content of this LG model, let us next construct a basis for the fluxes. We again start from the single model, in this case the $x^3$ LG model. The A-type D-branes that describe the integral homology are given by contours in the $x$-plane where $\mathcal{W}$ is of constant phase, which by an R-rotation can be taken to be as depicted in figure \ref{fig:contours2}. These wedges $V_0,V_1,V_2$ are linearly dependent,
\begin{equation}
    V_0+V_1+V_2=0\, .
\end{equation}
The $\mathbb{Z}_3$ symmetry which sends $g:x\mapsto \omega x$ acts on these wedges by exchanging them as
\begin{equation}
    g:V_{n}\mapsto V_{n+1} \,,
\end{equation}
with $n\in\mathbb{Z}_3$. The intersection between these wedges is given by $\langle V_m | V_n \rangle = \delta_{m,n}-\delta_{m+1,n}$, with $m,n \in \mathbb{Z}_3$. Having established this integral homology basis, let us next turn to the cohomology basis. For the single copy $x^3$ LG model, the cohomology basis is described by the chiral-chiral ring $\mathcal{R}=\mathbb{C}[x]$, which can be spanned by monomials $1,x$. These monomials correspond to R-R ground states $|\ell\rangle$ according to $\ell = k+1=1,2$. The intersection between this complex cohomology basis and the integral homology basis is computed through the contour integral
\begin{equation}
    \langle V_n | \ell \rangle = \int_{V_n}x^{\ell-1} e^{-\mathcal{W}} dx = \tfrac{1}{3}\omega^{n\ell}(1-\omega^{\ell})\Gamma\left(\tfrac{\ell}{3}\right)\, .
\end{equation}
For practical purposes it is convenient to adapt the normalization of these states to
\begin{equation}
    |\chi_{\ell} \rangle = \frac{3}{\Gamma\left(\tfrac{\ell}{3}\right)}|\ell \rangle
\end{equation}
We can then expand this complex basis in terms of the wedges as
\begin{equation}
    |\chi_{\ell} \rangle = \sum_n \omega^{n\ell}|V_{n}\rangle \, .
\end{equation}
While the basis $V_n$ is most convenient for the quantization, the advantage of the RR ground states $|\chi_{\ell}\rangle$ is that they provides an eigenbasis for the $\mathbb{Z}_3$ symmetry, since they transform under $g \in \mathbb{Z}_3$ as
\begin{equation}
    g:|\chi_\ell\rangle\mapsto \omega^{-\ell}|\chi_\ell\rangle \,.
\end{equation}
For completeness, let us record the overlap $\langle \chi_{\ell'} | \chi_{\ell} \rangle = 3 \delta_{\ell'+\ell,3}(1-\omega^\ell)$ between two RR ground states. We also note that complex conjugation relates them by $\overline{\chi_\ell} = \chi_{3-\ell}$.

\paragraph{Cohomology of the $(x^3)^{\otimes{9}}$ LG model.} With the knowledge of the single $x^3$ LG model, we now tensor 9 copies together to obtain the basis for the $(x^3)^{\otimes{9}}$ LG model. The R-R ground states are given by the tensor products
\begin{equation}\label{eq:tensor2}
    |\chi_{\bell}\rangle=|\chi_{\ell_1}\rangle\otimes |\chi_{\ell_2}\rangle \otimes...\otimes |\chi_{\ell_9}\rangle \,,
\end{equation}
where $\bell=(\ell_1,\ldots,\ell_9)$ with $\ell_i\in\{1,2\}$. The action of the $i$-th generator $g_i\in\mathbb{Z}_3^9$ on the RR ground states follows from the action on a single copy, which yields
\begin{equation}
    g_i:\chi_{\bell}\mapsto\omega^{-\ell_i}\chi_{\bell} \,.
\end{equation}
Invariant states that survive the orbifold projection $g=g_1\cdots g_9$ then correspond to RR charges that add up to
\begin{equation}
    \sum_{i=1}^{9}\ell_i\equiv 0\mod 3 \,.
\end{equation}
which is consistent with the conclusions from the chiral-chiral ring before. Similar to those deformations, there is a correspondence between the ground states with $\sum \ell_i=9,12,15,18$ and the Hodge decomposition into spaces $H^{3,0},H^{2,1},H^{1,2},H^{0,3}$, which has been summarized in table \ref{table:LGmodel2}. We can work out how the components of a three-form flux $H$ transform under the $\mathbb{Z}_3^9$ symmetry by expanding in this complex basis of RR ground states. Writing $H=\sum_{\bell}H^{\bell}\chi_{\bell}$, we find that its components transform as
\begin{equation}
    g_i:H^{\bell}\mapsto \omega^{-\ell_i}H^{\bell} \,.
\end{equation}
In order to work out the quantization of these fluxes, we also need the integral homology basis. This lattice is generated by the tensor product of $9$ A-type branes
\begin{equation}
    V_{\mathbf{n}}=V_{n_1}\otimes V_{n_2}\otimes...\otimes V_{n_9} \,,
\end{equation}
where $\mathbf{n}=(n_1,\ldots,n_9)$ with $n_i \in \{0,1,2\}$. The orbifold action requires us to restrict to the invariant sublattice $\gamma_{\mathbf{n}}$, which yields the equivalence relation\footnote{Here we again use a different but equivalent convention compared to \cite{Becker:2006ks, Becker:2023rqi, Becker:2024ijy}.}
\begin{equation}
    \gamma_{\mathbf{n}}\sim [V_{\mathbf{n}}]\sim [V_{\mathbf{n+1}}]\sim [V_{\mathbf{n+2}}] \,.
\end{equation}
While this basis is overcomplete, following \cite{Becker:2023rqi} a complete basis is obtained by considering the first 170 integers in base 2. Let us next turn to the intersections between these bases. The intersection pairing on the (orbifold) homology basis is given by
\begin{equation}
    \langle\gamma_{\mathbf{n'}}|\gamma_{\mathbf{n}}\rangle=\langle V_{\mathbf{n'}}|V_{\mathbf{n}}\rangle+\langle V_{\mathbf{n'+1}}|V_{\mathbf{n}}\rangle+\langle V_{\mathbf{n'+2}}|V_{\mathbf{n}}\rangle\,.
\end{equation}
Using this relation, the intersection between the cohomology and homology basis needed for the quantization condition of the fluxes is given by
\begin{equation}
    \langle \gamma_{\mathbf{n}}|\chi_{\bell}\rangle=3\omega^{\mathbf{n}\cdot \bell}\prod_{i=1}^{9}(1-\omega^{\bell_i}) \,,
\end{equation}
while the intersection pairing on the RR ground states reads
\begin{equation}
    \langle \chi_{\bell'}|\chi_{\bell}\rangle=\delta_{\bell'+\bell,\mathbf{3}}3^{10}\prod_{i=1}^{9}(1-\omega^{\bell_i}) \,.
\end{equation}

\begin{table}[]
    \centering
    \begin{tabular}{| c || c | c | c | c |}\hline
        $H^{p,q}$ & $H^{3,0}$ & $H^{2,1}$ & $H^{1,2}$ & $H^{0,3}$\\ \hline 
        $\sum_i \ell_i$ & 9 & 12 & 15 & 18 \\ \hline
        $\bell$ & $(1,1,1,1,1,1,1,1,1)$ &   $(1,1,1,1,1,1,2,2,2)$ &   $(1,1,1,2,2,2,2,2,2)$  & $(2,2,2,2,2,2,2,2,2)$ \\\hline
    \end{tabular}
    \caption{Summary of the Hodge decomposition into $(p,q)$-form spaces of the $(x^3)^{\otimes 9}$ LG model. We have listed $S_9$ representatives of the RR charges of the ground states \eqref{eq:tensor2} spanning these spaces.}
    \label{table:LGmodel2}
\end{table}

\paragraph{Orientifold.} Unlike the case of the $(x^4)^{\otimes 6}$ LG model, not all moduli and cycles in the $(x^3)^{\otimes 9}$ LG model are kept under the orientifold projection \eqref{eq:x3orientifold}. Namely, recall that O3/O7-orientifolds only keep the odd directions for the complex structure moduli and odd three-forms for the fluxes, cf.~\cite{Grimm:2004uq}. While all details for the various orientifolds of the $(x^3)^{\otimes 9}$ LG model can readily be found in \cite{Becker:2006ks}, let us briefly summarize how the projection \eqref{eq:x3orientifold} acts on the middle cohomology. The spaces $H^{3,0}$ and $H^{0,3}$ are spanned by states with $\bell=\mathbf{1},\mathbf{2}$, so these are odd and therefore kept under the orientifold projection as usual. For the space of $(2,1)$-forms $H^{2,1}$, i.e.~those ground states with RR charges $\sum_i \ell_i = 12$, it is convenient to separate into three cases:
\begin{itemize}
    \item $\ell_1=\ell_2=1$: the states $|\chi_{11\ell_3\cdots\ell_9}\rangle $ are odd, and therefore all 35 survive.
    \item $\{\ell_1,\ell_2\}=\{1,2\}$: the states $|\chi_{12\ell_3\cdots\ell_9}\rangle$ and $|\chi_{21\ell_3\cdots\ell_9}\rangle$ combine into odd and even combinations, and only the 21 odd combinations survive:
    \begin{equation}
       \frac{1}{\sqrt{2}}\big(|\chi_{12\ell_3\cdots \ell_9}\rangle + |\chi_{21\ell_3\cdots \ell_9}\rangle\big) \,.
    \end{equation}
    \item $\ell_1=\ell_2=2$: the states $|\chi_{22\ell_3\cdots\ell_9}\rangle$ are also odd, and hence all 7 survive.
\end{itemize}
Thus in total $h^{2,1}_-=63$ out of the initial $84$ $(2,1)$-forms survive the orientifold projection. The same (combinations of) complex structure moduli $t_{\mathbf{k}}$ survive, and all others are set to zero by the projection.

\paragraph{Flux superpotential.}
Similar to the $(x^4)^{\otimes 6}$ LG model, we can also express the classical spacetime superpotential into the contour integral. For a single component $G_3=\gamma_{\mathbf{n}}$, the superpotential is given by the deformed contour integral:
\begin{equation}
    W(\gamma_{\mathbf{n}},t_{\mathbf{k}})=\langle\gamma_{\mathbf{n}}|\mathbf{1}\rangle=\int_{\gamma_{\mathbf{n}}}d^6x e^{-\sum_{i=1}^{9}x_i^3+\sum_{\mathbf{k}}x^{\mathbf{k}}t_{\mathbf{k}}}\,.
\end{equation}
This integral can be used to compute the classical flux superpotential to all orders in the complex structure moduli $t_\mathbf{k}$. Similar to the $(x^4)^{\otimes 6}$ LG model, already from this contour integral form it is apparent how the symmetry $\mathbb{Z}_3^9$ acts on the superpotential: under the $i$-th generator $g_i$, the measure $d^9x$ picks up an factor of $\omega$, the worldsheet superpotential $\mathcal{W}(x_i;t_{\mathbf{k}})$ is invariant, and the contour is transformed by $\gamma_{\mathbf{n}}\mapsto\gamma_{\mathbf{n}+\mathbf{1}}$. Since the transformation of the contour is precisely how the flux $G_3$ transforms, we find that
\begin{equation}
    g_i\in\mathbb{Z}_3^9:\qquad W(G_3,t_{\mathbf{k}})\mapsto W(g_i\cdot G_3,\omega^{-k_i}t_{\mathbf{k}})=\omega W(G_3,t_{\mathbf{k}})\,.
\end{equation}
Hence, the spacetime flux superpotential again has charges $\mathbf{1}=(1,1,1,1,1,1,1,1,1)$, matching with the RR charges $\bell=\mathbf{1}$ of the holomorphic $(3,0)$-form.

Returning to the expansion of the flux superpotential, its all-order expression was readily derived in \cite{Becker:2024ijy}. For a single component $G_{3}=\chi_\bell$, the coefficient of the $t_{\mathbf{k_1}}\cdots t_{\mathbf{k_r}}$ is given by
\begin{equation}
    \partial_{t_{\mathbf{k_1}}}\cdots \partial_{t_{\mathbf{k_r}}} W(\chi_{\bell},t_{\mathbf{k}})|_{t_{\mathbf{k}}=0}=\delta(\mathbf{L}+\bell\mod 3)\prod_{i=1}^{9}(1-\omega^{L_i})\Gamma\left(\tfrac{L_i}{3}\right)\,,
\end{equation}
with $\mathbf{L}=\sum_{i=1}^{r}\mathbf{k}_i+\mathbf{1}$. Just like the $(x^4)^{\otimes 6}$ LG model, since this expression computes the periods of the $(3,0)$-form, it may also be used to compute the classical Kähler potential and scalar potential.

\section{Symmetries of the superpotential and scalar potential}\label{sec:symmetries}
In this section we use symmetries to constrain the expansion of the superpotential $W$, K\"ahler potential $K$, and scalar potential $V$ in terms of the moduli and the fluxes. This allows us to recast extremization conditions such as $W=\partial_i W=0$ and $\partial_i V=0$ as constraints on the flux components.

\subsection{Symmetries of the superpotential}
We begin by analyzing the superpotential. We first give a domain wall argument to explain why the superpotential is linear in the fluxes. We also describe how superpotential transforms under the relevant $\text{SL}(2,\mathbb{Z})$ symmetries and symmetries of the LG models. We then use these symmetries to constrain the expansion of the superpotential by selection rules based on the charges of the moduli and the fluxes.

\paragraph{Domain walls and flux superpotentials.} Let us begin by arguing why the superpotential is at most linear in the fluxes. To this end, we consider the domain wall associated to the flux \cite{Gukov:1999ya}. As we move from one side to the other, the flux value jumps from $G_3=0$ to some non-vanishing value $G_3$. The BPS tension $T$ of this domain wall is given by the change in value of the superpotential $W$ as we move across it
\begin{equation}
    T = |\Delta W|\, .
\end{equation}
We know that the tension $T$ of the domain wall is additive and thus linear in the fluxes since we can consider a sequence of them to accomplish the change in flux, so the change $\Delta W$ is linear as well. On the other hand, this only predicts how $W$ shifts, so we need to include a moduli-dependent constant $c \in \mathbb{C}$ as
\begin{equation}\label{eq:Wterms}
    W(F_3, H_3; t_{\mathbf{k}},\tau) = W_{\rm flux}(F_3,H_3; t_{\mathbf{k}}, \tau) + c(t_{\mathbf{k}},\tau)\, ,
\end{equation}
where $W_{\rm flux}$ is linear in the three-form fluxes $F_3, H_3$. Indeed, we know that non-perturbative corrections can in principle generate such corrections $c(t_{\mathbf{k}},\tau)$ to the flux superpotential \cite{Witten:1996bn}.  However one should also keep in mind that since the overall $W$ is a section of a line bundle, both the $W_{flux}$ term and the $c$-term should transform the same way under modular transformations.  In particular, since the first term will pick up a $1/(c\tau+d)$ under $\text{SL}(2,\mathbb{Z})$ modular transformations, so should the second term.  This parallels the argument we gave before that modular symmetries should be consistent with the instanton corrections in the geometric case.

\paragraph{Charges of fluxes.} In order to write down the selection rules for the superpotential, it is convenient to expand the fluxes in an eigenbasis of the symmetries \eqref{eq:SL2ZW} and \eqref{eq:Z4Z3W} of the superpotential. For the symmetries $\mathbb{Z}_4,\mathbb{Z}_3 \subset \text{SL}(2,\mathbb{Z})$ we can define the following linear combinations for the $(x^4)^{\otimes 6}$ and $(x^3)^{\otimes 9}$ LG models respectively
\begin{equation}
    H^{\pm } = F_3 \mp  i H_3\, \qquad H^{\pm } = F_3 \mp \omega H_3\, .
\end{equation}
Under the transformation rule \eqref{eq:SL2ZF3H3} we find that these combinations transform as
\begin{equation}
    S=\begin{pmatrix}
        0 & 1 \\
        -1 & 0
    \end{pmatrix} \in \mathbb{Z}_4: \quad H^{\pm } \mapsto i^{\pm 1} H^{\pm }\, , \quad U^2 =\begin{pmatrix}
        0 & 1 \\
        -1 & -1
    \end{pmatrix}\in \mathbb{Z}_3: \quad H^{\pm } \mapsto \omega^{\pm 1} H^{\pm }\, .
\end{equation}
Subsequently, we may decompose these into eigencomponents under the LG symmetries as described in \eqref{eq:Hexpand}. Then the flux quanta $H^{\pm, \bell}$ transform under the $i$-th generator of the LG symmetries as
\begin{equation}
    g_i \in \mathbb{Z}_4^6: \quad H^{\pm, \bell} \mapsto i^{-\ell_i} H^{\pm, \bell}\, ,\, \qquad     g_i \in \mathbb{Z}_3^6: \quad H^{\pm, \bell} \mapsto \omega^{-\ell_i} H^{\pm, \bell}\, .
\end{equation}
We want to point out that corrections to the 4d $\mathcal{N}=1$ supergravity theory cannot affect the transformation rules of these fluxes. The reason is that the symmetries are discrete phases and cannot continuously vary while representing a discrete group action.

\paragraph{Symmetries of the coordinates.} In order to make the discrete gauge symmetries manifest, we choose a canonical coordinate for the axio-dilaton following \cite{Bershadsky:1993cx}. It is provided by the holomorphic flat coordinate around $\tau=a$ given by
\begin{equation}\label{eq:canonicalcoor}
    \delta\tau_a(\tau) = \frac{\tau-a}{\tau - \bar{a}}\, ,
\end{equation}
which maps the upper-half plane into the unit disk. Its inverse is given by $\tau(\delta\tau) = (a-\bar{a} \delta\tau )/(1-\delta\tau)$. At the points of interest $a=i,\omega$, it is a straightforward exercise to verify that the symmetries $\mathbb{Z}_{3},\mathbb{Z}_4 \subset \text{SL}(2,\mathbb{Z})$ act on these coordinates through phase rotations
\begin{equation}\label{eq:coorsym}
    S \in \mathbb{Z}_4 : \quad  \delta\tau_i \mapsto -\delta\tau_i\, , \qquad \mathbb{Z}_3 :\quad \delta\tau_\omega \mapsto \omega^{-1} \delta\tau_\omega \, .
\end{equation}
The complex structure moduli $t_{\mathbf{k}}$ are neutral under SL$(2,\mathbb{Z})$ transformations, and instead transform under the $\mathbb{Z}_4^6$ and $\mathbb{Z}_3^9$ symmetries of the LG models as readily given in \eqref{eq:gtx46} and \eqref{eq:gtx39}.

\paragraph{Symmetries of the superpotential.} We next consider the action these discrete symmetries on the superpotential. We begin with $\mathbb{Z}_3,\mathbb{Z}_4 \subset \text{SL}(2,\mathbb{Z})$. In order to make these discrete gauge symmetries manifest, we have to make a different choice of K\"ahler gauge
\begin{equation}
    W \to (1-\delta\tau_a) W \, ,
\end{equation}
where $\delta\tau_{a}$ corresponds to the canonical coordinate introduced in \eqref{eq:canonicalcoor} for the fixed points $a=i,\omega$. The finite order elements of SL$(2,\mathbb{Z})$ then act on the superpotential at $\tau=i,\omega$ as
\begin{equation}\label{eq:SL2ZW}
    S \in \mathbb{Z}_4 : \quad W \mapsto i W\, ,\qquad  U^2\in \mathbb{Z}_3: \quad W \mapsto \omega W\, ,.
\end{equation}
This can be verified from the action of these symmetries on the axio-dilaton \eqref{eq:coorsym} and the RR and NS-NS fluxes at these points using the GVW superpotential \eqref{eq:W}.\footnote{In particular, because of the different K\"ahler gauge, $W$ also picks up these phases away from the elliptic points $a=i,\omega$, instead of the modular transformation \eqref{eq:SL(2,Z)W} in the original gauge.} We allow for corrections to this classical flux superpotential, but these should organize themselves into suitable modular objects as described in \eqref{eq:Wgen} such that the symmetry transformations of the superpotential do not change. 

Let us look more closely at the flux-independent term $c(t_{\mathbf{k}},\tau)$ in \eqref{eq:Wterms}. It should transform the same way as the flux-dependent terms above, otherwise modular symmetry would be broken, which as we have argued in section \ref{sec:dualitysymmetries} should be preserved.
But since the complex structure moduli are neutral under the modular group, and there are no other fields which could have in principle contributed to the weight in the flux-independent term, this will imply that the term without flux actually vanishes. This is because at $\tau =i$ the transformation of $\delta \tau \mapsto -\delta \tau $ cannot generate a phase $i$ for the superpotential, and there are also no other fields that transform. This argument for $c(t_{\mathbf{k}},\tau)=0$ also reflects the absence of modular forms of odd weight for SL$(2,\mathbb{Z})$, since the superpotential has weight $-1$.\footnote{It is crucial here that fluxes are present, since the flux superpotential transforms without any additional phases. In contrast, purely non-perturbative superpotentials as considered e.g.~in \cite{Font:1990gx,Cvetic:1991qm} can pick up phases under modular transformations, in which case functions transforming with automorphic factors---such as $\eta(\tau)$---would be allowed.}

Let us next turn to the symmetries $\mathbb{Z}_4^6$ and $\mathbb{Z}_3^9$ defined in \eqref{eq:Z4^6} and \eqref{eq:Z3^9} of the LG models. We know from the review of these LG models in section \ref{ssec:x46} and \ref{ssec:x39} that the holomorphic $(3,0)$-form transforms as a particular eigenstate with charges $\bell=\mathbf{1}$, and that the flux superpotential inherits the same charges through the standard formula \eqref{eq:W}. The resulting transformation rules read
\begin{equation}\label{eq:Z4Z3W}
    g_i \in \mathbb{Z}_4^6: \quad W \mapsto i W\, , \qquad g_i \in \mathbb{Z}_3^6 : \quad W \mapsto \omega W\, ,
\end{equation}
where $g_i$ denotes the generator of the $i$-th $\mathbb{Z}_4$ or $\mathbb{Z}_3$ respectively. Similar to the finite-order symmetries in SL$(2,\mathbb{Z})$, this holds even when corrections to the superpotential are included.

\paragraph{Selection rules.} With these preparations in place, let us then turn to the expansion of the superpotential in terms of the fluxes and the moduli. Readily enforcing the terms to transform accordingly under $\mathbb{Z}_3,\mathbb{Z}_4 \subset \text{SL}(2,\mathbb{Z})$, the general expansion of the superpotential reads
\begin{equation}\label{eq:Wexpansion}
\begin{aligned}
    W = &\sum_\bell c^+_{\bell} H^{+\bell} + \sum_{\bell} c^-_{\bell} H^{-\bell}\delta \tau + \sum_{\bell,\mathbf{k}} c^+_{\bell, \mathbf{k}} H^{+\bell} t_\mathbf{k} + \sum_{\bell,\mathbf{k}} c^-_{\bell, \mathbf{k}} H^{-\bell} \delta\tau t_\mathbf{k}\\
    & + \sum_{\bell,\mathbf{k},\mathbf{k}'} c^+_{\bell, \mathbf{k}, \mathbf{k}'} H^{+\bell}  t_\mathbf{k} t_{\mathbf{k}'}+\sum_\bell \tilde{c}^+_{\bell} H^{+\bell} \delta\tau^2 +\mathcal{O}(\delta\tau^3,\delta\tau^2 t_{\mathbf{k}},\delta\tau t_{\mathbf{k}}^2,t_{\mathbf{k}}^3)\, ,
\end{aligned}
\end{equation}
where we suppressed terms of cubic order or higher in the moduli fields. These higher-order terms can also be constrained by selection rules, but we have chosen to leave them out on purpose, since they are not relevant to the extremization conditions $D_I W = 0$ or $W=\partial_i W=0$. There are only flux-dependent terms, since we argued from SL$(2,\mathbb{Z})$ symmetry below \eqref{eq:Z4Z3W} that all flux-independent terms must vanish. Moreover, for the $(x^3)^{\otimes 9}$ LG model we note that the quadratic term in $\delta\tau^2$ should vanish $\tilde{c}^+_{\bell} = 0$ because of the $\mathbb{Z}_3 \subset \text{SL}(2,\mathbb{Z})$ symmetry, while in the $(x^4)^{\otimes  6}$ LG model it may be present. To get the full picture, let us now impose the $\mathbb{Z}_4^6$ and $\mathbb{Z}_3^9$ symmetries of the LG model, from which we derive the following selection rules
\begin{equation}\label{eq:Wselection}
\begin{aligned}
    c^{\pm}_{\bell} &= 0 \text{ unless $\bell=\mathbf{3}$ (or $\bell=\mathbf{2}$)}\, , \\
    \tilde{c}^+_{\bell} &= 0 \text{ unless $\bell=\mathbf{3}$}\, , \\
    c^{\pm}_{\bell,\mathbf{k}} &=0 \text{ unless $\bell = \mathbf{3}-\mathbf{k}$ (or $\bell = \mathbf{2}-\mathbf{k}$)}\, ,\\
    c^{\pm}_{\bell,\mathbf{k}, \mathbf{k}'} &=0 \text{ unless $\bell = \mathbf{3}-\mathbf{k}-\mathbf{k}'$ (or $\bell = \mathbf{2}-\mathbf{k}-\mathbf{k}'$)}\, .
\end{aligned}
\end{equation}
We gave the selection rules for the $(x^4)^{\otimes 6}$ LG model, and included the selection rules for the $(x^3)^{\otimes 9}$ LG model in between parentheses; as noted before, we have $\tilde{c}^+_{\bell} = 0$ in the latter LG model irrespective of the value of $\bell$. While we leave a detailed discussion of critical points of the scalar potential to section \ref{sec:criticalpoints}, let us make some preliminary remarks here. The vanishing of $c^+_{\bell}$ tells us whether the vacuum superpotential $W$ vanishes or not at the LG point $t_\mathbf{k}=0$, and at $\delta \tau=0$. Assuming it vanishes (or that the K\"ahler potential does not have any linear terms), the coefficient $c^-_{\bell}$ tells us whether the F-term along $\delta\tau$ vanishes, while the coefficients $c^+_{\bell,\mathbf{k}}$ indicate whether the F-terms along the complex structure moduli $t_{\mathbf{k}}$ vanishes. Finally, the quadratic terms tell us how many moduli can become massive, and here $c^-_{\bell,\mathbf{k}}$ tells us whether we get terms of the form $\delta\tau t_{\mathbf{k}}$, while $c^+_{\bell,\mathbf{k}, \mathbf{k}'}$ does so for $t_{\mathbf{k}}t_{\mathbf{k}'}$.

\subsection{Symmetries of the scalar potential}\label{ssec:sympotential}
We next apply these symmetry constraints to the expansion of functions that are neutral with respect to the duality symmetries. While the scalar potential $V$ which is neutral under symmetries is the main object of interest, the K\"ahler potential $K$ is also a neutral function, and so the same results apply there as well. The selection rules we derive are most effective at constant and linear order in the moduli, while for quadratic and higher-order many terms are generically expected to be present.

\paragraph{Expansion of scalar potential.} While we previously used a domain wall argument to explain why the superpotential is at most linear in the fluxes, we cannot use this argument to bound the number of fluxes in the expansion of the scalar potential. Namely, the K\"ahler potential is known to receive corrections due to fluxes in 4d $\mathcal{N}=1$ supergravity theories. We therefore consider arbitrary products of fluxes in the expansion of the scalar potential
\begin{equation}\label{eq:Vexpansion}
    V = \sum_{n_i, \tilde{n}_i, m_k, \tilde{m}_k} c_{\mathbf{n}\mathbf{\tilde n}\mathbf{m}\mathbf{\tilde m}} \bigg[ \prod_{\bell } (H^{+,\bell})^{n_{\bell} } (H^{-,\bar{\bell}})^{\tilde{n}_{\bell} } \bigg] \bigg[\prod_{\mathbf{k}}(t_{\mathbf{k}})^{m_{\mathbf{k}}}(\overline{t}_{\mathbf{k}})^{\tilde{m}_{\mathbf{k}}} \bigg] (\delta\tau)^{m_0}(\overline{\delta\tau})^{\tilde{m}_0}\, ,
\end{equation}
where the vectors $\mathbf{m}, \mathbf{\tilde m}$ include both the exponents $m_0,\tilde{m}_0$ of the axio-dilaton and the exponents $m_{\mathbf{k}},\tilde{m}_{\mathbf{k}}$ of the complex structure moduli.

\paragraph{Selection rules.} Demanding this scalar potential to be invariant under the  $\mathbb{Z}_4^7$ and $\mathbb{Z}_3^{10}$ coming from SL$(2,\mathbb{Z})$ and the LG model, we find as selection rules for the non-vanishing terms
\begin{equation}\label{eq:rule1}
    c_{\mathbf{n}\mathbf{\tilde n}\mathbf{m}\mathbf{\tilde m}} = 0 \quad \text{unless} \quad \begin{cases} \sum_{\bell}(n_\bell-\tilde{n}_\bell) &\equiv -2(m_0-\tilde m_0) \mod d \, ,\\
    \sum_{\bell} (n_\bell-\tilde{n}_\bell)\bell &\equiv -\sum_{\mathbf{k}}(m_{\mathbf{k}}-\tilde m_{\mathbf{k}}) \mathbf{k} \mod d\, ,
    \end{cases}
\end{equation}
where $d=3,4$ for the $(x^3)^{\otimes 9}$ and $(x^4)^{\otimes 6}$ LG model respectively. The first condition follows from imposing invariance with respect to $\mathbb{Z}_4, \mathbb{Z}_3 \subset SL(2,\mathbb{Z})$, while the second condition comes from the LG  symmetries and holds componentwise for the charges $\ell_i,k_i$. While at first sight this may look like an infinite set of conditions, notice that we only have finitely many values to check for the differences 
\begin{equation}
    \Delta n_\bell=n_\bell-\tilde{n}_\bell=0,\ldots, d \, , \qquad \Delta m_k=m_k-\tilde{m}_k= 0,\ldots, d-1\, ,
\end{equation}
where we used the short-hand notation $k=(0,\mathbf{k})$. As mentioned before, these selection rules can be applied to any function that is invariant with respect to the symmetry transformations, so apart from the scalar potential $V$ it also applies to the K\"ahler potential $K$.

\paragraph{Genericity of quadratic terms.} Let us now apply these selection rules to study what kind of quadratic terms in the moduli appear generically in the scalar potential. For simplicity we consider a critical point that has a non-vanishing cosmological constant $V \neq 0$. In this case, any quadratic correction to the K\"ahler potential automatically gives rise to a quadratic term in the scalar potential through the factor of $e^K$ in \eqref{eq:Vgen}.\footnote{In contrast, for Minkowski vacua we have $V=0$, so the vanishing of $D_I W$ and $W$ turns such quadratic corrections to the K\"ahler potential into higher-order terms in the scalar potential.} For a given pair of conjugate flux $H^{+,\ell},H^{-,\bar\ell}$ we can then consider the following quadratic terms in the scalar potential
\begin{equation}\label{eq:Vquadratic}
    V \supset H^{+,\bell}H^{-,\bar\bell} \,\delta\tau \overline{\delta\tau}+\sum_{\mathbf{k} } H^{+,\bell}H^{-,\bar\bell} \, t_{\mathbf{k}}\bar{t}_{\mathbf{k}} \, .
\end{equation}
where for simplicity we set the coefficients to 1. Clearly, all such terms are allowed by symmetries, since the phases picked up by fluxes and moduli cancel against their conjugates. Therefore, we expect generically that these critical points with $V \neq 0$ do fix all of the moduli. However, we cannot say anything about the sign of these quadratic terms, so we cannot comment on how many fields we expect to be massed up or tachyonic.

\section{Symmetry-protected critical points}\label{sec:criticalpoints}
Here we discuss supersymmetric Minkowski vacua, AdS vacua, and non-supersymmetric dS saddle points. We give the general criteria for the fluxes to realize such critical points based on the symmetry principles established in section \ref{sec:symmetries}. This is in the same spirit as what was done in \cite{Ginsparg:1986wr} for the non-supersymmetric $O(16)\times O(16)$ string in finding critical points where all the massless fields (except for the dilaton) are charged. We illustrate our findings by comparing to the results expected from the classical K\"ahler and superpotential, although since we generically expect corrections, the conclusions we draw for the exact masses from this numerical analysis should not be taken too seriously.

{\color{red} In the v1 and v2 of this paper we gave examples in the $(x^4)^{\otimes 6}$ LG model with $Q_{\rm flux}=40$. It was since then pointed out by \cite{Morros:2026yyi} that the correct normalization is $Q_{\rm flux}=20$. Wherever possible we have given new examples that saturate this updated tadpole bound.}

\subsection{Minkowski vacua}
In this section we study Minkowski vacua at the LG point for the complex structure moduli and the corresponding symmetric elliptic points for the axio-dilaton. We explain from symmetry principles what types of fluxes yield such Minkowski solutions, without the need to rely on the explicit form of the superpotential or invoke any non-renormalization theorems. We also argue from genericity of the quadratic terms that the vacua of \cite{Becker:2024ayh} remarkably stabilize all moduli using only the symmetries as our guide.\footnote{\color{red} The reference \cite{Becker:2024ayh} still used the old tadpole bound $Q_{\rm flux}=40$, and we have not been able to find new Minkowski examples that saturate the updated tadpole bound $Q_{\rm flux}=20$.}

\paragraph{Extremization conditions.} Let us first recall the geometric extremization conditions for Minkowski vacua. Demanding both the superpotential and the F-terms to vanish corresponds to $W=\partial_i W = 0$. In terms of the Hodge decomposition this amounts to requiring the complex three-form flux to lie in
\begin{equation}
    G_3 \in H^{2,1}\, .
\end{equation}
Let us now compare this condition with what we get from demanding $W=\partial_i W = 0$ by the selection rules for the expansion \eqref{eq:Wexpansion}. Let us denote by $H_{\ell}$ the eigenspace spanned by flux quanta $H^\bell$ with $\sum_i \ell_i = \ell$. Demanding the vacuum superpotential to vanish corresponds to
\begin{equation}\label{eq:W=0}
    W = 0 \quad \iff \quad H^{+} \in \begin{cases}
        H_{6}\oplus H_{10} \oplus H_{14} \quad &\text{ in the $(x^4)^{\otimes 6}$ LG model}\, ,\\
        H_{9}\oplus H_{12} \oplus H_{15} \quad &\text{ in the $(x^3)^{\otimes 9}$ LG model}\, ,
    \end{cases}
\end{equation}
as a piece along $H_{18}$ in the respective LG models would induce a non-vanishing vacuum superpotential. Geometrically this is precisely the piece of $G_3$ along the anti-holomorphic $(0,3)$-form spanning $H^{0,3}$. Demanding linear terms in the complex structure moduli also to vanish restricts these fluxes further to
\begin{equation}
    W = \partial_{t_\mathbf{k}} W = 0 \quad \iff \quad H^{+} \in \begin{cases}
        H_{6}\oplus H_{10} \quad &\text{ in the $(x^4)^{\otimes 6}$ LG model}\, ,\\
        H_{9}\oplus H_{12}  \quad &\text{ in the $(x^3)^{\otimes 9}$ LG model}\, ,
    \end{cases}
\end{equation}
On the other hand, requiring linear terms in the axio-dilaton to be absent constrains the flux $H^-$ as
\begin{equation}
    \partial_\tau W = 0 \quad \iff \quad H^- \in \begin{cases}
        H_{6}\oplus H_{10} \oplus H_{14} \quad &\text{ in the $(x^4)^{\otimes 6}$ LG model}\, ,\\
        H_{9}\oplus H_{12} \oplus H_{15} \quad &\text{ in the $(x^3)^{\otimes 9}$ LG model}\, ,
    \end{cases}
\end{equation}
as a piece along $H_{18}$ in the respective LG models would have introduced a term that is constant in the complex structure moduli $t_{\mathbf{k}}$ but linear in the axio-dilaton $\tau$. Finally, we should take into account that the fluxes are real, as conjugation relates $H^{+,\bell}$ to $H^{-,\bar\bell}$. In total, this yields the following necessary and sufficient conditions for a Minkowski vacuum at LG point $t_{\mathbf{k}}=0$ and the corresponding elliptic point for $\tau$:
\begin{equation}\label{eq:Minkowski}
\begin{aligned}
    \text{$(x^4)^{\otimes 6}$ LG model:}\qquad H^+ \in H_{10}\, , \ H^- \in  H_{14}\, ,\\
    \text{$(x^3)^{\otimes 9}$ LG model:}\qquad H^+ \in H_{12} \, , \ H^- \in H_{15}\, .
\end{aligned}
\end{equation}
These conditions correspond precisely to the requirement that $G_3 = H^+ \in H^{2,1}$ for Minkowski vacua, cf.~table \ref{table:LGmodel}. The symmetry selection rules for the expansion of $W$ guarantee that all these vacua survive corrections to the superpotential, without the need to assume any non-renormalization properties of the superpotential! In particular, the M-theory-like vacua of \cite{Becker:2024ayh} that mass up all moduli satisfy \eqref{eq:Minkowski}, so purely from symmetry considerations we have argued that for the existence of these Minkowski vacua!

\paragraph{Genericity of quadratic terms.} It is expected in quantum gravity that all types of terms and corrections are expected to appear, unless there is an underlying (super)symmetry principle that explains why they vanish \cite{Heckman:2019bzm, Palti:2020qlc}. To establish from symmetry principles whether all moduli are actually massed up, let us therefore look at what quadratic terms we generically expect to arise from the selection rules. We focus our attention on the Minkowski vacua presented in \cite{Becker:2024ayh} that masses up all moduli in the $(x^4)^{\otimes 6}$ LG model. We recall the selection rule \eqref{eq:Wselection} for the quadratic terms in the moduli
\begin{equation}\label{eq:quadratic}
    \begin{aligned}
            \partial_{t_{\mathbf{k}}} \partial_{t_{\mathbf{k}'}} W = 0 \quad \text{unless} \quad &H^{+,\bell} \neq 0  \ \ \text{ for }\bell = \mathbf{3}-\mathbf{k}-\mathbf{k}'\, ,\\
         \partial_{t_{\mathbf{k}}} \partial_{\delta\tau} W  =  0 \quad \text{unless} \quad &H^{-,\bell} \neq 0  \ \ \text{ for }\bell = \mathbf{3}-\mathbf{k}\, .\\   
    \end{aligned}
\end{equation}
The vanishing of $\partial_{\delta\tau}^2W = 0$ is guaranteed by the fact that $H^{+,\mathbf{3}} = 0$ for Minkowski vacua. The rank of the matrix $\partial_i \partial_J W$ tells us how many moduli will be stabilized. As a test case, we consider the matrix
\begin{equation}\label{eq:choice}
    \partial_I \partial_J W = \begin{cases}
        1 \text{ if allowed by \eqref{eq:quadratic},}\\
        0 \text{ otherwise\, .}
    \end{cases}
\end{equation}
For the example of \cite{Becker:2024ayh}, this gives a matrix with $91\times 91=8281$ entries, out of which 7460 vanish by symmetry considerations, which exactly matches with the number of vanishing entries when using the classical flux superpotential used in \cite{Becker:2024ayh}. Nevertheless, even for the relatively non-generic choice \eqref{eq:choice}, we find that $\partial_i \partial_J W$ has maximal rank, despite the fact that many of the diagonal entries vanish. Thus just from symmetry and genericity considerations we have found that these Minkowski vacua must exist and have all moduli massed up!

\subsubsection{Minkowski examples of \texorpdfstring{$(x^3)^{\otimes 9}$}{x39} LG model}\label{ssec:dSsaddlex39}
Here we study a couple examples of Minkowski vacua in the $(x^3)^{\otimes 9}$ LG model, from symmetry point of view.  We take three non-isolated Minkowski models constructed in \cite{Becker:2023rqi}. We already established in \eqref{eq:Minkowski} that all Minkowski vacua correspond to $H^+ \in H_{12}=H^{2,1}$. We show that we can deduce the number of massless modes from symmetry considerations and genericity, up to a quadratic term for $\delta\tau^2$ that may stabilize one more modulus. It would also be interesting to do a systematic search for isolated Minkowski and dS solutions for this LG model based on symmetry principles, which we will not consider further in this paper.

\paragraph{Genericity of quadratic terms.} Similar to the $(x^4)^{\otimes 6}$ LG model, we can first write down the selection rules for the mass matrix associated to the superpotential. We recall the selection rule \eqref{eq:Wselection} for the quadratic terms in the moduli
\begin{equation}\label{eq:quadratic2}
    \begin{aligned}
            \partial_{t_{\mathbf{k}}} \partial_{t_{\mathbf{k}'}} W = 0 \quad \text{unless} \quad &H^{+,\bell} \neq 0  \ \ \text{ for }\bell = \mathbf{2}-\mathbf{k}-\mathbf{k}'\, ,\\
         \partial_{t_{\mathbf{k}}} \partial_{\delta\tau} W  =  0 \quad \text{unless} \quad &H^{-,\bell} \neq 0  \ \ \text{ for }\bell = \mathbf{2}-\mathbf{k}\, .\\   
    \end{aligned}
\end{equation}
The vanishing of $\partial_{\delta\tau}^2 W = 0$ holds for any fluxes in the $(x^3)^{\otimes 9}$ LG model. Similar to \eqref{eq:choice}, we make the choice to set $\partial_I \partial_J W = 1$ whenever it should be non-vanishing by symmetries. In the following we now go through three examples constructed in \cite{Becker:2023rqi} with four components $H^{+,\bell}$ turned on and maximal tadpole $Q_{\rm flux} = 12$. These are characteristic of the three possible numbers 16, 22, and 26 of stabilized moduli with such fluxes.

The first case we consider is given by the flux
\begin{equation}
    H^+ = 3^{-7}  i \sqrt{3} \big(-\chi_{111111222}+\chi_{111112122}+\chi_{111121212}-\chi_{111122112}\big)\, .
\end{equation}
Using the classical superpotential, we find that these fluxes stabilize 16 moduli, and $4032$ out of the $64 \times 64=4096$ entries of $\partial_I \partial_J W$ vanish. From symmetry considerations we recover the same number of non-vanishing components, and find that the rank of the mass matrix is generically also 16.

The next example we consider is given by
\begin{equation}
    H^+ = 3^{-7}  i \sqrt{3} \big(-\chi_{111111222}+\chi_{111112122}+\chi_{111221211}-\chi_{111222111}\big)\, .
\end{equation}
Using the classical superpotential, these fluxes stabilize 22 moduli, and $4032$ entries of $\partial_I \partial_J W$ vanish as before. Symmetry considerations lead to the same conclusions, and also give $4032$ vanishing entries and generically a rank of 22.

As final example we consider the flux
\begin{equation}
    H^+ = 3^{-7} i \sqrt{3} \big(-\chi_{111111222}+\chi_{112111221}+\chi_{221111112}-\chi_{222111111}\big)\, .
\end{equation}
Using the classical superpotential, these fluxes stabilize 26 moduli, and now $4020$ entries of $\partial_I \partial_J W$ vanish. From symmetry considerations we also find $4020$ vanishing entries and generic rank $26$ for the mass matrix.

\subsection{de Sitter saddles}
Here we summarize our findings for dS saddle points in the LG models.  We will be looking for fluxes such that at the symmetric points have $W=0$, but $\partial W \not=0$, which guarantees $V>0$. We also look for $\partial_i V=0$ at the symmetric point. Using the selection rule \eqref{eq:rule1} we derived from symmetry principles, we argue that all examples we find are critical in all directions of the scalar fields: $\partial_i V=0$. Concretely, we classify all possible three-form fluxes with two or three components $H^{+,\bell}$ turned on. We select only solutions that satisfy the selection rules \eqref{eq:rule1}. {\color{red} We also demand the tadpole bound---which was updated from $Q_{\rm flux}=40$ to $Q_{\rm flux}=20$ in \cite{Morros:2026yyi} for the $(x^4)^{\otimes 6}$ LG model---to be saturated in these models, so that we do not have to worry about any moduli coming from mobile D3-branes.}

\paragraph{Extremization conditions.} Let us begin by studying the dS saddles from the perspective of the selection rules on the superpotential and scalar potential. In order to ensure that the cosmological constant $V$ is positive at the vacuum, we consider only fluxes with a vanishing vacuum superpotential $W=0$; only the F-terms can then be non-vanishing in \eqref{eq:Vgen}, and this contribution is manifestly positive. Geometrically this corresponds to
\begin{equation}\label{eq:dS}
    G_3 \in H^{3,0} \oplus H^{2,1} \oplus H^{1,2}\,  .
\end{equation}
From the point of view of the selection rules, demanding a vanishing vacuum superpotential $W=0$ corresponds to the same condition \eqref{eq:W=0} as previously encountered; for completeness, let us recall it here
\begin{equation}
\begin{aligned}
    \text{$(x^4)^{\otimes 6}$ model:}\quad H^{+} \in H_{6}\oplus H_{10} \oplus H_{14}\, ,\quad 
    \text{$(x^3)^{\otimes 9}$ model:}\quad H^{+} \in H_{9} \oplus H_{12}\oplus H_{15} \, ,
\end{aligned}
\end{equation}
which is the symmetry condition analogue to the geometric condition \eqref{eq:dS}, cf.~table \ref{table:LGmodel}. However, we still need to ensure that the scalar potential is extremized at the LG point with respect to all moduli. By specializing the selection rule \eqref{eq:rule1} to linear terms in the moduli, we obtain for derivatives along the complex structure moduli
\begin{equation}\label{eq:dVselect}
    \partial_{t_{\mathbf{k}}} V = 0 \quad \text{unless} \quad \begin{cases} \sum_{\bell}\Delta n_{\bell} &= 0 \mod d\, , \\
    \sum_{\bell}\Delta n_{\bell} \bell  &= - \mathbf{k} \mod d \quad \text{for some $t_{\mathbf{k}}$}\, , \\
    \end{cases}
\end{equation}
and for the derivative along the axio-dilaton
\begin{equation}\label{eq:dVselect2}
    \partial_\tau V = 0 \quad \text{unless} \quad \begin{cases} \sum_{\bell}\Delta n_{\bell} &= -2 \mod d\, , \\
    \sum_{\bell}\Delta n_{\bell} \bell  &= 0 \mod d \, , \\
    \end{cases}
\end{equation}
where $\Delta n_{\mathbf{\ell}}=0,1,2,3$ denotes the difference in factors of fluxes $H^{+,\bell}\neq 0$ and $H^{-,\bar\bell}\neq 0$ in the expansion of the scalar potential \eqref{eq:Vexpansion}. For a given flux $H^{\pm}$ with components $H^{\pm, \bell}\neq 0$, it is straightforward to check whether all linear terms vanish in the scalar potential according to this rule by going through all possible values of the $\Delta n_{\bell}$. We have performed a systematic analysis of the landscapes of dS saddle points in sections \ref{ssec:dSsaddlex46} and \ref{ssec:dSsaddlex39}, covering all possible fluxes with at most three components $H^{+,\bell}$ turned on. 

It is instructive to take a closer look at the requirement in this work to saturate the tadpole bound, i.e.~$Q_{\rm flux} = Q_{\rm O3}$. This requires us to turn on at least one component $H^{+,\bell} \in H^{2,1}$, since only such components contribute positively to the tadpole bound. For the $(x^4)^{\otimes 6}$ and $(x^3)^{\otimes 9}$ LG models, this amounts to at least one non-vanishing component in $H^{+,\bell} \in H_{10}$ or $H^{+,\bell} \in H_{12}$ respectively. In this case, the selection rule \eqref{eq:dVselect} does not allow for any flux along $H_{6}$ or $ H_{9}$ in the respective LG models, so we are left with considering fluxes in the following eigenspaces
\begin{equation}
        \text{$(x^4)^{\otimes 6}$ model:}\quad H^{\pm} \in  H_{10} \oplus H_{14}\, ,\quad 
    \text{$(x^3)^{\otimes 9}$ model:}\quad H^{\pm} \in H_{12} \oplus H_{15}\, ,
\end{equation}
which, in geometric terms, corresponds to fluxes of the type $H^+ \in H^{2,1}\oplus H^{1,2}$. For the examples we construct in subsection \ref{ssec:dSsaddlex46} we take these considerations into account, but otherwise we allow for all types of dS saddles in our discussion, not necessarily restricting to those saturating the tadpole bound.

\paragraph{Finiteness.} For ordinary flux vacua with vanishing F-terms the finiteness of the number of solutions $D_IW=0$ follows from imposing the tadpole bound and dividing out by duality groups. In fact, it has been proven \cite{Bakker:2021uqw, Grimm:2021vpn} that for a given Calabi--Yau fourfold the landscape of these flux vacua is finite (see also \cite{Grimm:2023lrf, Grimm:2024fip}). For the dS saddles considered here this does not apply, since we have non-vanishing F-terms, so there are also negative contributions. Geometrically these negative contributions come from the components of $G_3$ along $H^{3,0}$ and $H^{1,2}$ that turn on the F-terms $D_\tau W$ and $D_{t_{\mathbf{k}}}W$ respectively. For definiteness, let us consider the $(x^4)^{\otimes 6}$ LG model, and note that the $(x^3)^{\otimes 9}$ LG model follows analogously. The fluxes $H^{+,\bell} \in H_{10}$ contribute positively to the tadpole for the  (or $(x^3)^{\otimes 9}$ LG model), but the fluxes $H^{+,\bell} \in H_{6}\oplus H_{14}$ which cause $D_I W\neq 0$ contribute negatively. This can be seen explicitly from the tadpole contribution
\begin{equation}
    Q_{\rm flux} = \frac{4^5}{2i} \sum_{\bell}|H^{+,\bell}|^2\prod_{i=1}^6 (1-i^{-\ell_i})  \, ,
\end{equation}
where we note that the product $\prod_{i=1}^6 (1-i^{-\ell_i})$ is purely imaginary with a positive imaginary part for $H^{+,\bell} \in H_{10}$ and negative imaginary part for $H^{+,\bell} \in H_{6}\oplus H_{14}$. Nevertheless, we can still argue for a finite landscape by requiring the cosmological constant $V$ to lie within the regime of validity of the EFT. To be precise,  we should bound the Hubble scale $H = \sqrt{V}$ of the dS saddle point to lie below the cut-off $\Lambda_{\rm EFT}$ of the EFT \cite{Hebecker:2018vxz,Scalisi:2018eaz,vandeHeisteeg:2023uxj}, i.e.~$V \leq \Lambda_{\rm EFT}^2$. For our dS saddle points the vacuum superpotential vanishes $W=0$, so this cosmological constant is made up entirely of the F-terms $D_IW\neq 0$, so we get
\begin{equation}\label{eq:VLambda}
    V = e^K K^{I\bar J}D_I W D_{\bar J}\bar{W} \Big|_{t_{\mathbf{k}}=0,\delta\tau=0} \leq \Lambda_{\rm EFT}^2\, .
\end{equation}
Recall from the selection rules for the superpotential \eqref{eq:Wselection} that only the fluxes with $H^{+,\bell} \in H_6 \oplus H_{14}$ can contribute to the linear part of $W$, and thereby to the constant part of $D_IW\neq0$ and thus the cosmological constant $V$. At the same time, the left-hand side of \eqref{eq:VLambda} furnishes a positive definite norm on the space $H_{6}\oplus H_{14}$ of these states, which geometrically corresponds to the norm on the imaginary anti-selfdual part of the three-form cohomology. As we do not expect corrections to cause this norm to become degenerate, this means that there are also only finitely many flux choices in $H_{14}$ until we violate the cut-off bound \eqref{eq:VLambda}. 

\begin{table}[]
    \centering
    \begin{tabular}{c || c | c | c| c | c | c | c | c  }
$\text{eig}(K^{IK}\partial_i \partial_J V) /V$ & -5.1100 & -2 & 0 &1 & $1.0151$ & $\frac{5}{4}$ & $1.6660$ & $\frac{9}{4}$  \\ \hline 
mult. & 1 & 2 & 82 & 22 & 2 & 52 & 1 & 12 
    \end{tabular}\hspace{0pt}
    \begin{tabular}{ c | c | c | c | c |}
     $5.2317$ & 18.806 & 19.157 & 21.642 & 39.825 \\ \hline
     1 & 2 & 2 & 2 & 1
    \end{tabular}\hspace{0pt}
    \caption{Classical mass spectrum of the scalar potential for the dS example $H^+=\frac{7-i}{32}\chi_{1,1,2,2,2,2}+\frac{3i-1}{16}\chi_{3,3,1,1,3,3}$ with $V_E=500$. We evaluated the masses in units of the dS cosmological constant $V$ and also listed their multiplicities.  }
    \label{table:dSmasses}
\end{table}

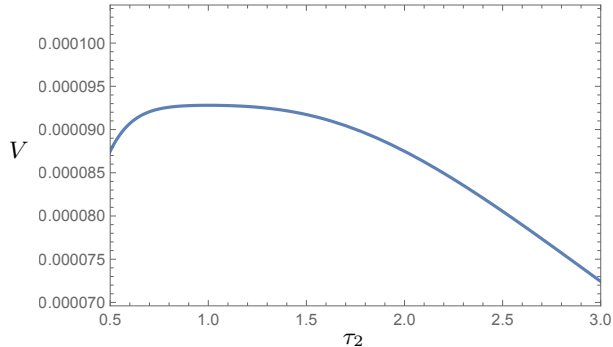
\begin{figure}
    \centering
    \includegraphics[width=0.5\linewidth]{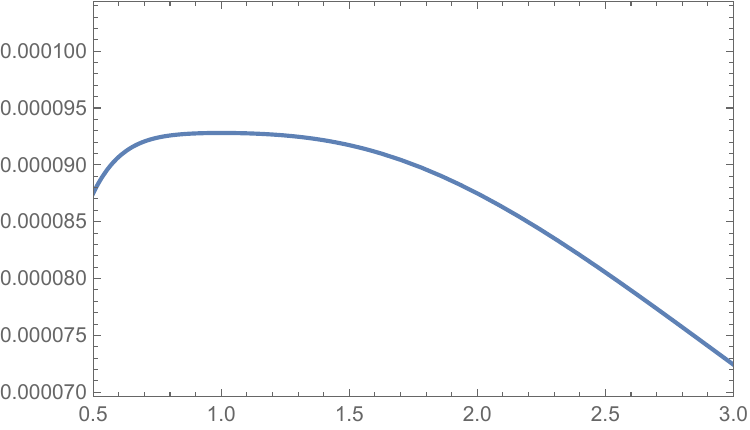}
    \begin{picture}(0,0)\vspace*{-1.2cm}
    \put(-230,65){\footnotesize $V$}
    \put(-105,-5){\footnotesize $\tau_2$}
    \end{picture}\vspace*{-0.1cm}
    \caption{Plot of the classical scalar potential \eqref{eq:Vclassical} for the flux $H^+=\frac{7-i}{32}\chi_{1,1,2,2,2,2}+\frac{3i-1}{16}\chi_{3,3,1,1,3,3}$ along the direction $\tau_2$ with $\tau_1=0$ and all complex structure moduli $t_{\mathbf{k}}=0$. We set $V_E=500$.}
    \label{fig:dS}
\end{figure}

\paragraph{Tachyonic directions.} All of the critical dS points we have obtained already possess tachyonic directions at the classical level. For an example we have included the values in table \ref{table:dSmasses}. The (refined) dS conjecture \cite{Obied:2018sgi,Garg:2018reu,Ooguri:2018wrx} proposes that whenever $\partial_i V = 0$,\footnote{Detailed investigations of the gradient $|\nabla V|/V$ have been performed for asymptotic potentials in string compactifications, see e.g.~\cite{Grimm:2019ixq, Calderon-Infante:2022nxb, Andriot:2025gyr}. Systematic analyses of tachyons in dS solutions coming from Type II supergravities have been performed in \cite{Andriot:2022bnb, Andriot:2024cct}.} the smallest eigenvalue satisfies
\begin{equation}
    \min\{\text{eig}(K^{IK}\partial_K \partial_J V) \}\leq - c' V\, ,
\end{equation}
for some $\mathcal{O}(1)$ constant $c'$ in Planck units.  We cannot rigorously check this condition, because we do not know the exact K\"ahler potential in particular.  However, we can use the naive classical formula to see whether in that approximation it is being satisfied. We included the metric factor $K^{IK}$ here to canonically normalize the scalar fields. Among our examples, in the classical approximation we find $c'=-2.28514$ as the smallest value, and thus at least in this approximation the refined dS conjecture is satisfied. 

Let us next take a closer look at where they come from. While we do not have a general argument for tachyons, based on the examples we find that there are two ways in which they can arise. In most cases we find at least one tachyon to come from the dilaton, like depicted in figure \ref{fig:dS}. Let us focus on this particular example, corresponding to the last row in table \ref{table:twochis}, since it is also the model with the lowest number of tachyons we have found. Setting the complex structure moduli $t_{\mathbf{k}}=0$, the classical scalar potential for the axio-dilaton reads
\begin{equation}\label{eq:Vclassical}
    V_{\rm classical} \big|_{\tau_1=t_{\mathbf{k}}=0} = e^{K_{\rm a-d}(\tau,\bar\tau)}\big(320 - 40 \tau_2 + 180 |\tau|^2 \big)\, .
\end{equation}
The coefficients here depend on the flux quanta, and a different choice can lead to a different structure of this scalar potential. Setting $\tau_1=0$ and expanding the K\"ahler potential \eqref{eq:Kadcs} around $\tau_2 = 1$ we find
\begin{equation}
    \partial_{\tau_2}^2 V_{\rm classical} \big|_{\tau=t_{\mathbf{k}}=0} = 360\frac{V_E + 90  E_{\frac{3}{2}}(i,-i) -80 \partial_{\tau_2}^2E_{\frac{3}{2}}(i,-i) }{\left(V_E + 90 E_{\frac{3}{2}}(i,-i) \right)^3}
\end{equation}
This gives rise to a tachyon provided the Einstein-frame volume $V_E$ is not too big (which we cannot determine even in the classical approximation); the precise numerical value corresponds to
\begin{equation}
    V_E \lesssim  588.431\, .
\end{equation}
This is to be contrasted with the minimal value that $V_E$ has to take for a positive-definite metric at $\tau=i$, which corresponds to $V_E \gtrsim  406.513$. Nevertheless, we note that this classical tachyon is not universal among all examples, since we have other examples where $\tau_2$ has a positive mass-squared for all $V_E \gtrsim  406.513$; for example, the second example in table \ref{table:twochis}.

Another way in which we have found tachyons to appear in some examples is through the direction $t_{\mathbf{k}'}$ of the F-term $D_{t_{\mathbf{k}'}}W \neq 0$ that we turned on.\footnote{Even though this seems to work as a good principle for two-flux examples, the three-flux examples in table \ref{table:threechis} provide counter-examples, since the tachyons come from other directions in the complex structure moduli. } Namely, in the same example as we just considered, we find that the F-term along $t_{\mathbf{k}'} = t_{002200}$ does not vanish, and indeed we find a quadratic term with a negative coefficient
\begin{equation}
    V \supset -\frac{4.17600}{\left(V_E + 813.03\right)^2} |t_{002200}|^2\, .
\end{equation}
Also notice that this tachyon cannot be removed by making $V_E$ arbitrarily large, in contrast to the dilaton.

\paragraph{Protection against runaway directions.} All of the examples found in this work possess many flat directions. Ordinarily, this would not be favorable for dS points, as any corrections to the scalar potential could turn these massless moduli into runaway directions instead. This is indeed what happens in known examples~\cite{Ginsparg:1986wr,Fraiman:2023cpa,Baykara:2024tjr}, where the string coupling furnishes a runaway direction for the one-loop cosmological constant. The difference with our case is that we have a symmetry argument that protects the scalar potential from acquiring such runaway directions, since all of our fluxes obey the selection rule \eqref{eq:rule1}, and therefore no linear terms in the moduli can be generated by any sort of correction, since it needs to fit with the duality symmetry. Nevertheless, as discussed below \eqref{eq:Vquadratic}, corrections quadratic in the moduli are expected to appear generically, so we do expect these directions to either become massed up or tachyonic.

\subsubsection{dS saddle landscape of \texorpdfstring{$(x^4)^{\otimes 6}$}{x46} LG model}\label{ssec:dSsaddlex46}
Using the selection rule \eqref{eq:dVselect}, we now set out to perform a systematic analysis of the dS saddle landscape of the $(x^4)^{\otimes 6}$ LG model. We go up to three fluxes $H^{+,\bell}$ and include only those where we can saturate the tadpole bound. This rules out any dS saddles with a single flux $H^{+,\bell}$, since we cannot simultaneously break the F-terms and have a positive tadpole contribution.

\begin{table}[]
    \centering
    \begin{tabular}{c || c | c| c|}
       $H^+$  & $n_{\rm massive}$ & $n_{\rm massless}$ & $n_{\rm tachyon}$ \\ \hline
$\frac{3+i}{32} \chi_{111223} + \frac{1+3i}{64}\chi_{222332}$ & 116 & 46 & 20 \\
$\frac{3+i}{32} \chi_{111223} + \frac{1+3i}{64}\chi_{223232}$ & 103 & 62 & 17 \\
$    \frac{7-i}{32} \chi_{112222} + \frac{3i-1}{16}\chi_{131333}$ & 108 & 59 & 15 \\
$\frac{7-i}{32} \chi_{112222} + \frac{3i-1}{16}\chi_{331133}$ & 97 & 82 & 3 \\
    \end{tabular}
    \caption{Two-$\chi$ fluxes of the form \eqref{eq:threechi} that satisfy the selection rules \eqref{eq:dVselect}. {\color{red} These examples have tadpole charge $Q_{\rm flux}=40$, double what is allowed by \cite{Morros:2026yyi}; no two-$chi$ examples with $Q_{\rm flux}=20$ exist.} The determination of spectrum of masses is done by using the classical approximation, which is expected to change when corrections are taken into account.  In particular the symmetries imply that the massless states are expected to shift and become either massive or tachyonic, but not runaway.}
    \label{table:twochis}
\end{table}

\paragraph{Two-$\chi$ solutions.} Next we move to examples with two fluxes $H^{+,\bell},H^{+,\bell'}$, which are of the form
\begin{equation}
    H^+ = H^{+,\bell_{10}} + H^{+,\bell_{14}}\, ,
\end{equation}
where $H^{+,\ell_{10}} \in H_{10}$ and $H^{+,\ell_{14}} \in H_{14}$. We need a flux along $\bell_{14}$ to have a positive $V>0$ at the vacuum. This flux by itself would have a negative tadpole contribution, so we compensate by adding a flux along $\bell_{10}$ that has a positive contribution. We enumerated all possible such pairs and found 13 combinations. {\color{red} For 4 of those we were able to find a solution with $Q_{\rm flux}=40$, but none with $Q_{\rm flux}=20$; we have nevertheless kept these examples for their illustrative purpose, } summarized in table \ref{table:twochis}.

\paragraph{Three-$\chi$ solutions.} We next collect all examples with three distinct fluxes $H^{+,\bell}$. The most general flux with three components that can saturate the tadpole bound reads
\begin{equation}\label{eq:threechi}
    H^+ = H^{+,\bell_{10}} +H^{+,\bell_{10}'} + H^{+,\bell_{14}}\, , \quad \text{or} \, \quad  H^+ = H^{+,\bell_{10}} +H^{+,\bell_{14}} + H^{+,\bell_{14}'}\, ,
\end{equation}
since we need at least one flux in $H_{10}$ to get a positive tadpole, and one along $H_{14}$ to generate a non-vanishing F-term. We enumerated all possible such triples compatible with the selection rules and found 28 combinations. {\color{red} For 2 of those we were able to find a solution with $Q_{\rm flux}=20$.} These have been summarized in table \ref{table:threechis}.

\begin{table}[]
\color{red}
    \centering
    \begin{tabular}{|c || c | c| c|}\hline 
       $H^+$  & $n_{\rm massive}$ & $n_{\rm massless}$ & $n_{\rm tachyon}$ \\ \hline $\left(\frac{1}{32}-\frac{i}{32}\right) \chi _{111313}-\left(\frac{1}{64}-\frac{i}{32}\right) \chi
   _{222112}-\left(\frac{1}{32}+\frac{i}{32}\right) \chi _{333311}$ & 131 & 29 & 22 \\
   $\left(-\frac{1}{64}+\frac{i}{16}\right) \chi _{112222}+\left(\frac{1}{32}-\frac{i}{32}\right) \chi
   _{331133}+\left(\frac{1}{32}-\frac{3 i}{32}\right) \chi _{333311}$ & 119 & 56 & 7 \\ \hline

    \end{tabular}
    \caption{\color{red} dS saddle points from three-$\chi$ fluxes of the form \eqref{eq:threechi} that satisfy the selection rules \eqref{eq:dVselect} and saturate the tadpole bound $Q_{\rm flux}=20$ of \cite{Morros:2026yyi}.}
    \label{table:threechis}
\end{table}

\subsection{AdS vacua}
We now turn to supersymmetric AdS vacua. We restrict our attention to vacua with the moduli stabilized at the respective elliptic points $\tau=i,\omega$ of the LG models. We study these vacua both from the perspective of the classical K\"ahler potential \eqref{eq:Kadcs} and superpotential \eqref{eq:W} and from general symmetry considerations.

\paragraph{Extremization conditions.} Let us begin by analyzing the classical extremization conditions. The Eisenstein series $E_{\frac{3}{2}}(\tau,\bar\tau)$ in $K_{\rm a-d}(\tau,\bar\tau)$ is extremized at the symmetric elliptic points, so this piece does not contribute to the F-term equations. Therefore we are left with the condition
\begin{equation}
    D_{t_{\mathbf{k}}} W  \int H^+ \wedge D_{t_{\mathbf{k}}} \Omega = 0\, \qquad D_\tau W  = \frac{1}{2i}\int H^- \wedge \Omega = 0\, .
\end{equation}
This corresponds to the usual restriction of the Hodge type
\begin{equation}
    H^+ \in H^{(2,1)}\oplus H^{(0,3)}\, ,
\end{equation}
where the distinction between a Minkowski or AdS vacua comes down to whether the piece along $H^{(0,3)}$ vanishes or not. It is helpful to compare this condition with the Hodge types found in \cite{Becker:2007dn, Becker:2022hse}. Away from the elliptic point for $\tau$ the Eisenstein series $E_{\frac{3}{2}}(\tau,\bar\tau)$ contributes to $\partial_\tau K$ and thereby selects a different direction in the plane $H^{(3,0)}\oplus H^{(0,3)}$. In the weak-coupling limit this contribution is $3/(\bar\tau-\tau)$, which requires this component of $H^+$ to lie along $-3\Omega+\bar\Omega$; this condition was readily observed in \cite{Becker:2007dn, Becker:2022hse}. 

We can arrive at similar conclusions for AdS vacua at the elliptic points from just symmetry considerations. We need to assume that the fluxes are chosen according to the selection rules \eqref{eq:dVselect} such that the K\"ahler potential is extremized. This assumption allows us to ignore the pieces $\partial_{t_{\mathbf{k}}} K=\partial_{\tau}K=0$ in the F-term equations, which now just amount to demanding the absence of linear terms in the superpotential
\begin{equation}
    D_{t_{\mathbf{k}}}W = \partial_{t_{\mathbf{k}}} W = 0\, , \qquad D_{\tau}W = \partial_\tau W = 0\, .
\end{equation}
From the selection rules for the superpotential expansion \eqref{eq:Wexpansion} we then learn that
\begin{equation}
\begin{aligned}
    \text{$(x^4)^{\otimes 6}$ LG model:}\qquad H^+ \in H_{10}\oplus H_{18}\, , \ H^- \in H_{6} \oplus H_{14}\, ,\\
    \text{$(x^3)^{\otimes 9}$ LG model:}\qquad H^+ \in H_{12} \oplus H_{18}\, , \ H^- \in H_{9}\oplus H_{15}\, .
\end{aligned}
\end{equation}
We used here that conjugation relates $H^{+,\bell}$ to $H^{-,\bar\bell}$, which sets the component of $H^{+}$ along $H_6$ in the $(x^4)^{\otimes 6}$ LG model ($H_9$ in the $(x^3)^{\otimes 9}$ LG model to zero), and similarly $H^{-}$ along $H_{10}$ in the $(x^4)^{\otimes 6}$ LG model ($H_{12}$ in the $(x^3)^{\otimes 9}$ LG model). 

\begin{table}[]
    \centering
    \begin{tabular}{c || c | c | c| c | c | c | c | c | c }
$m^2 R_{\rm AdS}^2$ & -$\frac{9}{4}$ & -2.24373 & -2.24203 & -$2$ & -$1.40211$ & -$\frac{5}{4}$ & -$1.06345$ & $411.997$ &  $453.703$ \\ \hline 
mult. & 34 & 1 & 1 & 108 & 1 & 34 & 1 & 1 & 1
    \end{tabular}
    \caption{Classical mass spectrum for the AdS example \eqref{eq:HAdS} with $V_E=500$. We listed the masses in units of $R^2_{\rm AdS} = -3/V $ and gave their multiplicities. }
    \label{table:AdSmasses}
\end{table}

\begin{figure}
    \centering
    \includegraphics[width=0.5\linewidth]{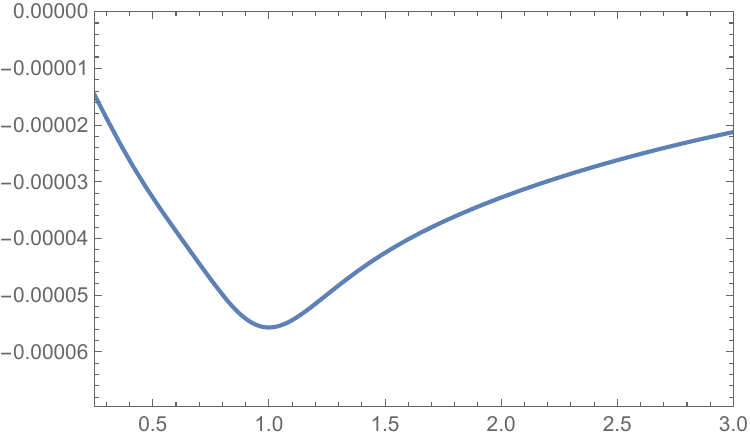}
    \begin{picture}(0,0)\vspace*{-1.2cm}
    \put(-230,65){\footnotesize $V$}
    \put(-105,-5){\footnotesize $\tau_2$}
    \end{picture}\vspace*{-0.1cm}
    \caption{Plot of the classical scalar potential for the flux given in \eqref{eq:HAdS} along the direction $\tau_2$ with $\tau_1=0$ and all complex structure moduli $t_{\mathbf{k}}=0$. We set $V_E=500$.}
    \label{fig:AdS}
\end{figure}
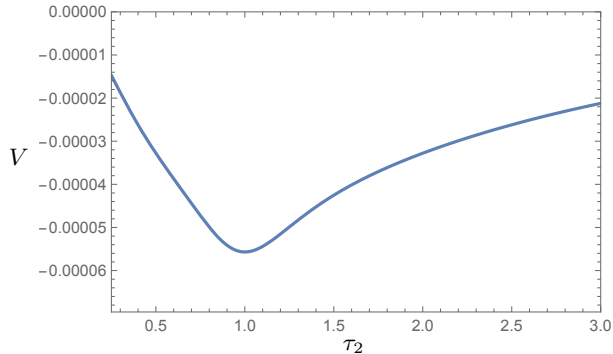

\paragraph{Example.} As an example, let us consider the following flux in the $(x^4)^{\otimes 6}$ LG model
\begin{equation}\label{eq:HAdS}
    H^+ = \left(-\frac{1}{64}-\frac{i}{64}\right) \left(\chi _{2,2,2,2,1,1}+4 \chi _{3,3,3,3,3,3}\right)\, .
\end{equation}
This example was already considered in \cite{Becker:2006ks}, and our symmetry argument now confirms the existence of these supersymmetric AdS vacua: this flux \eqref{eq:HAdS} does not permit any linear terms in the K\"ahler potential according to the selection rule \eqref{eq:rule1}, so it exactly satisfies $D_IW=0$. {\color{red} This example saturates the old tadpole bound with $Q_{\rm flux}= 40$.\footnote{\color{red} We have not been able to find new AdS vacua saturating the updated tadpole bound $Q_{\rm flux}=20$ of \cite{Morros:2026yyi}.}} It is instructive to look at the classical values for the cosmological constant and the masses. We set $V_E = 500$ to ensure a positive-definite K\"ahler metric. We included a plot of the classical scalar potential in figure \ref{fig:AdS}. The classical value of the AdS constant is then given by
\begin{equation}
    V_{\rm classical} \big|_{\rm t_{\mathbf{k}}=\delta\tau=0} = -\frac{96}{\left(V_E + 813.03\right)^2}\, ,
\end{equation}
which evaluates to $V_{\rm classical}=-0.000041762$ for $V_E=500$. We also computed the Hessian for this classical scalar potential and determined the masses of the canonically normalized scalar fields, which have been listed in table \ref{table:AdSmasses}. As required by the 4d $\mathcal{N}=1$ supergravity formalism, see for example \cite{Plauschinn:2022ztd}, all masses satisfy the BF bound \cite{Breitenlohner:1982jf} $m^2 \geq 9R_{\rm AdS}^2/4$. In fact, 34 fields saturate the BF bound, another 146 scalars are also tachyonic, and only two scalars have a positive mass-squared. And while the numerical values we found should be taken with a grain of salt, also notice that most masses take rational values; in light of the AdS/CFT correspondence they correspond according to $\Delta(\Delta-3)=m^2R_{\rm AdS}^2$ to operators in the dual 3d CFT with (half-)integer conformal dimensions $\Delta=\tfrac{3}{2}$, $\Delta=1,2$, and $\Delta=\tfrac{1}{2},\tfrac{5}{2}$. 

\section{Concluding Thoughts}\label{sec:conclude}
In this paper we have argued that the existence of critical points in AdS, Minkowski and dS vacua can in some cases be anticipated by symmetry principles for all the scalar modes.  A prominent symmetry we argued for is the modular invariance of the type IIB coupling in the LG orientifolds with flux. We argued that all the scalars are charged under the combination of discrete LG symmetries and modular symmetries.

Even though it sounds a bit far-fetched, one potential caveat to this argument is that there may be additional light degrees of freedom at strong couplings that becomes light and perhaps is not critical at the strong coupling point where all the other scalar light fields are critical. It is hard to imagine how this could happen concretely.
Another possible loophole in the argument for the dS or AdS is that the value of $|\Lambda| >1$ in Planck units, in which case these vacua are not physical.  This would not affect the argument for the Minkowski vacua where $\Lambda=0$.
Also we have assumed genericity of values for all the terms which are not forced to vanish due to symmetries. It could be that this is not the case, and some terms have accidental zeros and some directions remain massless. This could in principle invalidate an isolated Minkowski vacuum. Of course this does not affect the statement that $dV/d\phi_i=0$ even for these directions.

In this paper we have argued for the existence of unstable dS critical points.  Finding a meta-stable dS would be difficult to prove using our techniques, because we cannot trust the exact values of masses near the critical points. But we could have ended up with a critical dS point which at the classical level has no tachyons.  We did not find such an example in the classication approximation and all the dS examples we found have tachyonic directions with sufficient rate of instability, consistent with the dS conjectures \cite{Obied:2018sgi,Garg:2018reu,Ooguri:2018wrx,Bedroya:2019snp}. One may ask whether an unstable dS could potentially be relevant for late-time cosmology.  Indeed it was argued in \cite{Agrawal:2018rcg} that a top of the hill potential is perfectly consistent with both the late-time cosmology as well as the dS conjectures.  One of the worries in such a scenario would be that the assumption that the field starts close to the top of the potential may be an unnatural fine-tuning.  However, if the top of the hill corresponds to an enhanced symmetry point, as is the case in our models, this may not pose a problem: Top of the hill potential with an enhanced gauge symmetry point corresponds to spontaneous breaking of the gauge symmetry as we roll away.  However,  the universe could have started at a higher temperature where the generic expectation is that the symmetries get restored getting rid of the tachyonic directions and the symmetric point emerging as a preferred  point.  As the universe cools off the symmetry gets broken and the tachyonic direction emerges leading to a natural rolling away from the symmetric point and the start of late-time cosmology.

\acknowledgments
We thank Rafael \'Alvarez-Garc\'ia, Sergio Cecotti, Thomas Grimm, Amineh Mohseni, Muthusamy Rajaguru, Timm Wrase, and Kai Xu for helpful discussions. 

This work is supported in part by a grant from the Simons Foundation (602883, CV) and the DellaPietra Foundation.

\bibliographystyle{JHEP}
\bibliography{refs}

\end{document}